\documentclass[prd,preprint,superscriptaddress]{revtex4}
\usepackage{revsymb}
\usepackage{amsmath}
\usepackage{graphicx}
\usepackage{bm}
\newcommand{\be}{\begin{equation}}
\newcommand{\ee}{\end{equation}}
\newcommand{\ben}{\begin{eqnarray}}
\newcommand{\een}{\end{eqnarray}}

\begin{document}
\title{Is the cosmological dark sector better modeled by a generalized Chaplygin gas or by a scalar field?}
\author{Sergio del Campo\footnote{\dag\ deceased}} \affiliation{Instituto de F\'{\i}sica,
Pontificia Universidad Cat\'{o}lica de Valpara\'{\i}so, Avenida
Brasil 2950, Casilla 4059, Valpara\'{\i}so, Chile}
\author{J\'{u}lio C. Fabris\footnote{E-mail: fabris@pq.cnpq.br}}
\affiliation{Universidade Federal do Esp\'{\i}rito Santo,
Departamento
de F\'{\i}sica\\
Av. Fernando Ferrari, 514, Campus de Goiabeiras, CEP 29075-910,
Vit\'oria, Esp\'{\i}rito Santo, Brazil}
\author{Ram\'{o}n Herrera\footnote{E-mail: ramon.herrera@ucv.cl}}
\affiliation{Instituto de F\'{\i}sica, Pontificia Universidad
Cat\'{o}lica de Valpara\'{\i}so, Avenida Brasil 2950, Casilla
4059, Valpara\'{\i}so, Chile}
\author{Winfried Zimdahl\footnote{E-mail: winfried.zimdahl@pq.cnpq.br}}
\affiliation{Universidade Federal do Esp\'{\i}rito Santo,
Departamento
de F\'{\i}sica\\
Av. Fernando Ferrari, 514, Campus de Goiabeiras, CEP 29075-910,
Vit\'oria, Esp\'{\i}rito Santo, Brazil}

\begin{abstract}
Both scalar fields and (generalized) Chaplygin gases have been widely used separately to characterize the dark sector of the Universe.
Here we investigate the cosmological background dynamics for a mixture of both these components and quantify the fractional abundances that are admitted by observational data from supernovae of type Ia and from the evolution of the Hubble rate.
Moreover, we study how the growth rate of (baryonic) matter perturbations is affected by the dark-sector perturbations.
\end{abstract}
\date{\today}
\maketitle

\section{Introduction}

The standard cosmological model, the $\Lambda$CDM model ($\Lambda$ denotes the cosmological constant, CDM stands for cold dark matter),  assumes the presently observed cosmic substratum mainly to consist of a cosmological-constant type dark-energy (DE) together with pressureless CDM. These dark components make up about 95\% of the cosmic energy budget. ``Usual", i.e. baryonic, matter only contributes with less than 5\%. The present fractions of radiation and curvature are dynamically negligible.
The standard model describes well a large number of observations and its parameters have been determined by now with
high precision \cite{planck}.
On the other hand, the theoretical status of the standard model is anything but satisfactory. The model relies on the existence of a dark sector which is physically not really understood. Moreover, despite of its observational success there remain tensions \cite{buchert15}.
While no straightforward fundamental progress seems to be in sight at the moment,
this situation requires further (semi-) phenomenological  studies of potential deviations from the standard model
as well as of modifications both of the matter sector (right-hand side of Einstein's equations) and of the geometric  sector (left-hand side of Einstein's equations).

The simple  cosmological-constant model has been ``dynamized" in several ways. Even before the advent of the
observations of supernovae of type Ia (SNIa) by \cite{Riess,Schmidt,Perlm} which supported the idea of a universe in accelerated expansion, a scalar field (SF) has been suggested as an agent that might drive the cosmological dynamics \cite{wetterich}.
Further studies along this line were performed in \cite{Ratra,CDS,CCQ,PeeVi,Zlatev,Steinh,Martin,Barreiro,Luca,ZPC}.

A fluid dynamical description which is able to account both for an early matter-dominated phase and for effects similar to those generated by a cosmological constant has been established in terms of (generalized) Chaplygin gases. The original Chaplygin gas \cite{Chaplygin} is characterized by an equation of state (EoS)
$p = - \frac{A}{\rho}$. It was applied to cosmology in  \cite{moschella} followed by \cite{julio,bilic}.
A phenomenological generalization to an  EoS
$p=-\dfrac{A}{\rho^{\alpha}}$
with a constant $\alpha>-1$ was introduced in \cite{berto}, where also its relation to a scalar-field Lagrangian of a generalized Born-Infeld type was clarified.
For $\alpha =1$ this generalization reduces to the original Chaplygin gas, for $\alpha =0$ it is related to the $\Lambda$CDM model.
An appealing feature of the (generalized) Chaplygin gas (GCG) is its capability of a unified description of the dark sector.
Its  energy density  is changing smoothly from that of nonrelativistic matter at high redshift to an almost constant far-future value.
Thus it interpolates between an early phase of decelerated
expansion, necessary for successful structure formation, and a late period in which
it acts similarly as a cosmological constant, generating an accelerated expansion.
Cosmological models relying on the dynamics of generalized Chaplygin gases have been widely studied  in the literature \cite{oliver08,oliver09,gcgjulio1,ioav,rrrr,VDF1,VDF2,gcgjulio2,nadgcg,wands,saulo14,luciano}.

While both SF based models and models aiming at a unified description of the dark sector of the type of
GCGs have separately attracted ample attention, our aim in this paper is to investigate a model in which a GCG  and a SF are simultaneously present (GCSF model) in addition to a pressureless matter component which is supposed to describe the baryon fraction of the Universe.
By suitable parameter choices the GCSF model has two $\Lambda$CDM limits  which allows us to investigate deviations from the latter in various directions. We use SNIa and $H(z)$ data to test whether the observations admit a dark sector of the GCSF type.
The SF dynamics will be described with the help of the CPL parametrization \cite{CPL}. While this may seen as a loss of generality, it has the advantage of providing us with an explicit analytic expression for the Hubble rate. The existence of an analytic solution of the background dynamics is essential for the perturbation analysis.
This solution determines the coefficients of the system of coupled first-order perturbation equations.

The best-fit values of the background analysis are then used for a study of the growth rate of the (baryonic) matter perturbations. With the help of a simplifying parametrization of the dark-sector perturbations we investigate the impact of the latter on the matter-perturbation growth.

In Sec.~\ref{basic} we recall the basic properties of the GCG  and
SF components of the dark sector and find the Hubble rate of the GCSF model. The background data analysis and its interpretation is the subject of Sec.~\ref{observations}. Section~\ref{perturbations} is devoted to the sub-horizon dynamics of matter perturbations,
while  Sec.~\ref{summary} summarizes our results.


\section{The cosmic substratum}
\label{basic}
\subsection{Cosmic medium as a whole}
We assume a perfect-fluid structure of the cosmic medium as a whole, described by the energy-momentum tensor
\begin{equation}
T_{ik} = \rho u_{i}u_{k} + p h_{ik}\ ,\qquad\ h^{ik} = g^{ik} + u^{i} u^{k}\,, \qquad T_{\ ;k}^{ik} = 0,
\label{T}
\end{equation}
where $\rho = T_{ik}u^{i}u^{k}$ is the total energy density, $p = \frac{1}{3}T_{ik}h^{ik}$ is the total  pressure and $u^{i}$ is the four-velocity of the cosmic substratum as a whole, normalized to $u^{i}u_{i} = -1$.

\subsection{Decomposition into 3 components}
The total energy-momentum tensor in (\ref{T}) is split into a GCG (subindex $c$), a SF component (subindex $s$) and a matter component (subindex $m$),
\begin{equation}\label{Ttot}
T^{ik} = T_{c}^{ik} + T_{s}^{ik} + T_{m}^{ik}.
\end{equation}
We assume perfect-fluid structures of each of the components as well
($A= c, s, m$) and separate energy-momentum conservation
\begin{equation}\label{TA}
T_{A}^{ik} = \rho_{A} u_A^{i} u^{k}_{A} + p_{A} h_{A}^{ik} \
,\qquad\ h_{A}^{ik} = g^{ik} + u_A^{i} u^{k}_{A}\,,\qquad T_{A\ ;k}^{ik} = 0,
\end{equation}
where $\rho_{A} = T_{A}^{ik}u_{Ai}u_{Ak}$.
In general, the $4$ velocities of the components are different from each other and from the total
four velocity $u^{i}$ as well.

\subsection{Equations of state}
The equation of state for the GCG  is
\begin{equation}\label{eosC}
p_{c} = - \frac{A}{\rho_{c}^{\alpha}}.
\end{equation}
A simple scalar field (quintessence) is characterized by an EoS parameter $\omega_{q}$,
\begin{equation}\label{eosq}
   \omega_{q} = \frac{\frac{1}{2}\dot{\phi}^{2} - V(\phi)}{\frac{1}{2}\dot{\phi}^{2} + V(\phi)}.
\end{equation}
This parameter is restricted to $-1\leq \omega_{q}\leq 1$.
Under more general circumstances, e.g., for non-minimally coupled scalar fields or scalar fields with a non-standard kinetic term a phantom-type EoS is possible as well.
For the SF we use the
effective fluid description
\begin{equation}\label{eosS}
p_{s} = \omega_{s}\rho_{s}
\end{equation}
which is supposed to cover both the quintessence and the phantom  cases.
As far as the matter component with
\begin{equation}\label{}
p_{m} = 0
\end{equation}
is concerned, our main interest here is baryonic matter, but in some special cases below  also CDM will be included.

\section{Background dynamics}
\subsection{Conservation equations}
For a a homogeneous, isotropic and spatially flat universe with a Robertson-Walker metric, the total
energy-momentum conservation in (\ref{T}) reduces to
\begin{equation}\label{}
\dot{\rho} + 3H \left(\rho + p\right) =0,
\end{equation}
where $H= \frac{\dot{a}}{a}$ is the Hubble rate and $a$  is the scale factor of the Robertson-Walker metric.
In this background all the four-velocities are assumed to coincide,
\begin{equation}\label{}
u_{c}^{a} = u_{s}^{a} = u_{m}^{a} = u^{a}.
\end{equation}
The energy conservation equations for the components are
\begin{equation}\label{}
\dot{\rho}_{A} + 3H \left(\rho_{A} + p_{A}\right) =0.
\end{equation}

\subsection{Energy densities}

\subsubsection{Chaplygin gas}
With the EoS (\ref{eosC}) we obtain the energy density $\rho_{c}$,
\begin{equation}\label{}
p_{c} = - \frac{A}{\rho_{c}^{\alpha}}\qquad \Rightarrow\qquad \rho_{c} = \left[A + Ba^{-3\left(1+ \alpha\right)}\right]^{\frac{1}{1+\alpha}},
\end{equation}
where $B$ is a non-negative constant,
or
\begin{equation}\label{}
\rho_{c} = \rho_{c0}\left[\bar{A} + \left(1 - \bar{A}\right)a^{-3\left(1+ \alpha\right)}\right]^{\frac{1}{1+\alpha}}\,,\qquad \bar{A} = \frac{A}{\rho_{c0}^{1+\alpha}},
\end{equation}
where $\rho_{c0}$ is the energy density for $a=1$.
An EoS parameter $\omega_{c}$ is introduced via
\begin{equation}\label{wC}
\omega_{c}\equiv \frac{p_{c}}{\rho_{c}} = -\bar{A}\left(\frac{\rho_{c0}}{\rho_{c}}\right)^{1+\alpha}
\end{equation}
and the adiabatic sound speed by
\begin{equation}\label{}
\frac{\dot{p}_{c}}{\dot{\rho}_{c}} = - \alpha \frac{p_{c}}{\rho_{c}}.
\end{equation}
For $\bar{A} = 0$ the GCG reduces to a pure matter component.

\subsubsection{Scalar field in CPL parametrization}
For the general EoS (\ref{eosC}) one has
\begin{equation}\label{}
p_{s} = \omega_{s}\rho_{s}\qquad \Rightarrow\qquad \rho_{s} = \rho_{s0}a^{-3}\exp{\left[-3\int \omega_{s}(a)\frac{da}{a}\right]},
\end{equation}
which can either be quintessence with (\ref{eosq}) or phantom matter with $\omega_{s} < -1$.
The adiabatic sound speed is
\begin{equation}\label{}
\frac{\dot{p}_{s}}{\dot{\rho}_{s}} = \omega_{s} - \frac{\dot{\omega}_{s}}{3H\left(1 + \omega_{s}\right)}.
\end{equation}
For the purpose of this paper we adopt
the frequently used CPL \cite{CPL}  parametrization $\omega_{s} = \omega_{0} + \omega_{1}\left(1-a\right)$ with the help of which we have
the explicit formula
\begin{equation}\label{}
\rho_{s} = \rho_{s0}a^{-3\left(1 + \omega_{0} + \omega_{1}\right)}\exp{\left[3\omega_{1}\left(a-1\right)\right]}.
\end{equation}
In this manner the scalar-field dynamics is reduced to a two-parameter fluid
description. Such approximation is expected to make sense close to the present
time, i.e., for small redshift.
The total EoS parameter of the cosmic substratum is
\begin{equation}\label{}
\omega = \frac{p}{\rho} = \frac{p_{c} + p_{s}}{\rho}.
\end{equation}

\subsection{The Hubble rate}
Friedmann's equation reads
\begin{equation}
3 H^{2} = 8\pi G \rho = 8\pi G \left(\rho_{c} + \rho_{s} + \rho_{m}\right).
\label{fried}
\end{equation}
Introducing the fractional quantities
\begin{equation}\label{}
\Omega_{c0} = \frac{8\pi G\rho_{c0}}{3H_{0}^{2}},\quad \Omega_{s0} = \frac{8\pi G\rho_{s0}}{3H_{0}^{2}},\quad \Omega_{m0} = \frac{8\pi G\rho_{m0}}{3H_{0}^{2}},
\end{equation}
the Hubble rate is given by
\begin{equation}\label{hubble}
\frac{H^{2}}{H^{2}_{0}} = \frac{\rho}{\rho_{0}} = \Omega_{C0}\left[\bar{A} + \left(1 - \bar{A}\right)a^{-3\left(1+ \alpha\right)}\right]^{\frac{1}{1+\alpha}}
+ \Omega_{S0}a^{-3\left(1+w_{0}+w_{1}\right)}
e^{3w_{1}\left(a-1\right)}+   \Omega_{B0}a^{-3}.
\end{equation}
With the explicit expression (\ref{hubble}) the background dynamics is analytically known.
The $\Lambda$CDM model is recovered both for $\alpha =0$ together with  $\rho_{s0} =0$ (no SF, the GCG accounts both for DE and CDM) and for $A=0$ together with $\omega_{0} =-1$ and $\omega_{1} =0$ (vanishing kinetic term of the SF,  the GCG accounts for CDM only).

\section{Observations and statistical analysis}
\label{observations}

Now we confront the CGSF model  with data from SNIa and $H(z)$ data.
The five free parameters of the model are $\alpha$, $\Omega_{s0}$, $\bar{A}$, $\omega_{1}$ and $h$, where $h$ is defined by $H_{0}=100h\ \mathrm{kms^{-1}Mpc^{-1}}$. The parameter $\omega_{0}$ will be fixed to either $\omega_{0} =-1.0$
or, alternatively, to $\omega_{0} =-1.05$ or $\omega_{0} =-0.95$.
To get a better understanding of the combined model, we shall also evaluate the limiting cases of a vanishing SF contribution (the GCG describes the entire dark sector) as well  as the case of a SF with a certain amount of pressureless matter (no GCG).

We use the binned  set of supernovae data from the JLA compilation \cite{JLA}.
This test relies on the observed distance modulus $\mu_{obs}(z)$ of each binned SN Ia data at some redshift $z$,
\begin{equation}\label{muth}
\mu_{th}(z)=25+5 log_{10}\frac{d_L(z)}{Mpc},
\end{equation}
where the luminosity distance $d_L$ in a spatially flat Robertson-Walker metric, is given by the formula
\begin{equation}
d_L(z)=c(1+z)\int^{z}_0\frac{dz^{\prime}}{H(z^{\prime})}.
\end{equation}
The binned JLA data set contains 31 data points. The corresponding $\chi^2$ function is constructed according to
\begin{equation}
\chi^{2}_{SN}= \left(\mathbf{\mu}_{th}(z)-\mu_{obs}(z)\right)^{\dagger} \mathbf{C}^{-1} \left(\mu_{th}(z)-\mu_{obs}(z)\right),
\end{equation}
where $\bf{C}$ is the covariance matrix \cite{JLA}.

As a second observational source we consider the evaluation of differential age data of old galaxies that have evolved passively \cite{Ji1, Ji2, Hz,Farooq,moresco}. Here we use the 36 measurements of $H(z)$ listed in \cite{Zheng} which consist of 30 differential age measurements and 6 data from an analysis of baryon acoustic oscillations (BAO).
The relevant relation here is
\begin{equation}
H(z)=-\frac{1}{1+z}\frac{dz}{dt}.
\end{equation}
The spectroscopic redshifts of galaxies are known with very high accuracy. A differential measurement of time $dt$ at a given redshift interval allows one to obtain values for $H(z)$.
The chi-square function for the analysis of the $H(z)$ data is
\begin{equation}
\chi^2_{H}=\sum^{N_{H}}_{i=1}\frac{\left(H^{th}(z_i)-H^{obs}(z_i)\right)^2}{\sigma^2_i},
\end{equation}
where $N_{H}$ is the number of data points
and $\sigma_i$ is the observational error associated to each observation $H^{obs}$ while $H^{th}$ is the theoretical value predicted by the GCSF model.

Combining the information from both tests, we construct the total chi-square function as
\begin{equation}\label{chitotal}
\chi^2_{Total}=\chi^2_{SN}+\chi^2_{H}.
\end{equation}
The one-dimensional probability distribution functions (PDF) are obtained from  the likelihood function
\begin{equation}
\mathcal{L}= A e^{ -\chi^2_{Total}(\alpha, \Omega_{s0}, \bar{A},\omega_{1}, h )/2}
\end{equation}
by suitable marginalization procedures.

The results of the statistical analysis for the most general  case with all five parameters left free are listed in the first three lines
of TABLE I. The matter part is fixed here to $\Omega_{m0} = 0.04$, i.e., purely baryonic matter.
The reduced Hubble rate, $\alpha$ and $\bar{A}$ are almost unaffected by a change in $\omega_{0}$.
The present EoS parameter of the GCG, which according to (\ref{wC}) coincides with minus $\bar{A}$, is close to zero, i.e., this component behaves like dust. The value of $\alpha$ is unexpectedly large. The fraction $\Omega_{s0}$ has a slight tendency to increase with increasing $\omega_{0}$ (decreasing $|\omega_{0}|$). The $\Omega_{s0}$ values are larger than the corresponding value $\Omega_{\Lambda0}$ of the $\Lambda$CDM model. The EoS parameter $\omega_{1}$ is positive and decreases with decreasing $|\omega_{0}|$.
The fourth and fifth lines of TABLE I describe GCG universe models without a SF contribution.
The value $\Omega_{m0}$ quantifies the matter contribution additional  to that part which is already taken into account
by the GCG itself. In the fourth line $\Omega_{m0} = 0.04$ is assumed as in the three previous lines.
By considering $\Omega_{m0} = 0.3$ (fifth line) we admit a larger additional  matter fraction. Strictly speaking, such choice is
against the motivation of a unified description of the dark sector, but it is included here for comparison.
In the fourth line we recover the conventional, well-known Chaplygin-gas dynamics.
The fifth line with a present EoS parameter $-0.059$ does not correspond to an accelerated expansion of the present Universe. There might have been accelerated expansion in the past, however.
The last two lines of TABLE I describe SF universes with $\omega_{0}=-1 $  without a CGC. In the second-last line the matter fraction $\Omega_{m0} = 0.04$ is purely baryonic again, i.e., the SF accounts for the entire dark sector, in the last line with $\Omega_{m0} = 0.3$ the Universe is made of a SF and CDM. This case is close to the $\Lambda$CDM model at the present time.
One may suspect that the large value of $\alpha$ and the small values of $\bar{A}$ (both not typical for viable Chaplygin gas models)
in the first three lines indicate degeneracies in the parameter space.
To get a consistent picture we performed an alternative analysis the results of which are summarized in  TABLE II.
The values  of $\alpha$, $\Omega_{s0}$, $\bar{A}$ and $\omega_{1}$  in  TABLE II are obtained by
fixing $h$ to its best-fit values $h\approx 0.7$, resulting from a marginalization over the remaining variables. Otherwise, the configurations are the same as in TABLE I. The differences in the parameter values of both tables are remarkable, those of TABLE II are closer to expectations for Chaplygin-gas cosmologies.
In TABLE II the SF fraction for the mixed model (first three lines) is somewhat lower than that of the corresponding  $\Lambda$CDM model (while it was somewhat higher than the latter in TABLE I).
According to this analysis a certain part of the GCG, characterized by an EoS parameter $\omega_{c0}= - \bar{A}\approx -0.536$
for $\omega_{0}=-1$,  has to contribute to the dark energy as well.
For $\omega_{0}=-1.05$ and $\omega_{0}=-0.95$ a similar statement holds with different values for $\bar{A}$.
At the same time the $\alpha$  are drastically reduced to values close to zero. These features enforce our belief that the results of the procedure leading to TABLE II is superior to the 5-parameter analysis of TABLE I.
For the GCG only (fourth line) we have a present EoS value of $-0.729$, almost coinciding with the corresponding value in TABLE I, i.e., the pure GCG dynamics is recovered again.
The results for the pure SF (last two lines) remain unaltered as well.

\begin{table}
\label{tab1}
\begin{tabular}{|l|c|c|c|c|c|c|}\hline
Model&$\chi^2_{min}$&$\alpha$&$\Omega_{s0}$&$\bar A$&$\omega_1$&$h$ \\ \hline
GCSF, $\omega_{0} = -1.05,\ \Omega_{m0} = 0.04$ &52.425&8.000&0.761&0.050&0.599&0.708\\ \hline
GCSF, $\omega_{0} = -1.00,\ \Omega_{m0} = 0.04$ &51.046&8.000&0.768&0.050&0.514&0.706\\ \hline
GCSF, $\omega_{0} = -0.95,\ \Omega_{m0} = 0.04$ &50.533&8.000&0.775&0.050&0.432&0.704\\ \hline
GCG, $\Omega_{m0} = 0.04$ &50.562&-0.166&0.000&0.728&-&0.701\\ \hline
GCG, $\Omega_{m0} = 0.30$ &56.933&-57.188&0.000&0.059&-&0.689\\ \hline
SF, $\omega_{0} = -1.00,\ \Omega_{m0} = 0.04$&71.383&-&0.96&-&1.475&0.720\\ \hline
SF, $\omega_{0} = -1.00,\ \Omega_{m0} = 0.30$&55.363&-&0.70&-&-0.510&0.702\\ \hline
\end{tabular}
\caption{Best-fit values for the 5 parameter CGSF model with $\omega_{0}$ and $\Omega_{m0}$ fixed (first three lines). Fourth line: pure GCG, no SF, fixed baryon content, fifth line: pure GCG, no SF, fixed additional matter content.  Second last line: pure SF, no GCG, fixed baryon fraction, last line: pure SF, no GCG, fixed matter fraction. }
\end{table}

\bigskip
\begin{table}
\label{tab2}
\begin{tabular}{|l|c|c|c|c|}\hline
Model&$\alpha$&$\Omega_{s0}$&$\bar A$&$\omega_1$ \\ \hline
GCSF, $\omega_{0} = -1.05,\ \Omega_{m0} = 0.04$&-0.079&0.632&0.412&0.571\\ \hline
GCSF, $\omega_{0} = -1.00,\ \Omega_{m0} = 0.04$&-0.006&0.653&0.536&0.284\\ \hline
GCSF, $\omega_{0} = -0.95,\ \Omega_{m0} = 0.04$&0.086&0.670&0.614&0.630\\ \hline
GCG, $\Omega_{m0} = 0.04$&-0.169&0.000&0.729&-\\ \hline
GCG $\Omega_{m0} = 0.30$&-32.247&0.000&0.063&-\\ \hline
SF, $\omega_{0} = -1.00,\ \Omega_{m0} = 0.04$&-&0.96&-&1.476\\ \hline
SF, $\omega_{0} = -1.00,\ \Omega_{m0} = 0.30$&-&0.70&-&-0.511\\ \hline
\end{tabular}
\caption{Best-fit values for the CGSF model. Configurations as in TABLE 1 but with $h$ fixed to its best-fit value.}
\end{table}

A robust picture is obtained in terms of the one-particle PDFs in FIGS. 1-5.
FIG.~1 shows the one-dimensional PDFs for $\alpha$, $\Omega_{s0}$, $\bar{A}$ and $\omega_{1}$ where $h$ was fixed to its corresponding best-fit value and we assumed $\omega_{0} = -1$.
The same analysis for $\omega_{0} = -0.95$ has been performed in FIG.~2 and for $\omega_{0} = -1.05$ in FIG.~3.
The limiting case that there is no SF and the dark sector is entirely modeled by the GCG is visualized in FIG.~4.
In Fig.~5 we show the one-dimensional PDFs for $\omega_{1}$ in the opposite limit in which the GCG is absent and we are left with a SF cosmology with matter fractions of $0.04$, i.e., the matter is entirely baryonic (left figure)
or a matter fraction of $0.3$ corresponding to the total matter fraction, largely given by  CDM.

A comparison of figures 1-3 indicates that a variation in $\omega_{0}$ in the vicinity of $\omega_{0} = -1$
does not substantially affect the results of the analysis.
The stable features are a value of $\alpha$ close to  zero, a present SF fraction slightly larger than $0.6$, a present EoS value for the GCG of the order of $\omega_{c0} = - \bar{A}\approx - 0.6$
and a positive value smaller than (but of the order of) one for $\omega_{1}$.
The circumstance that $\Omega_{s0}\gtrsim 0.6$ seems to indicate that the background data prefer a SF dominated dark sector over a GCG dominated configuration.

In the following section we use the best-fit values listed in the second line of  TABLE II to get insight into the behavior of matter perturbations. Although done in a simplified manner this analysis is expected to capture
the essential features of the perturbation dynamics.
Notice that for the chosen configuration the matter perturbations are perturbations of the baryonic matter. After all, it is the baryonic matter distribution which is observed in galaxy catalogues.

\begin{figure}
\label{1}
\centering
\includegraphics[width=.225\textwidth]{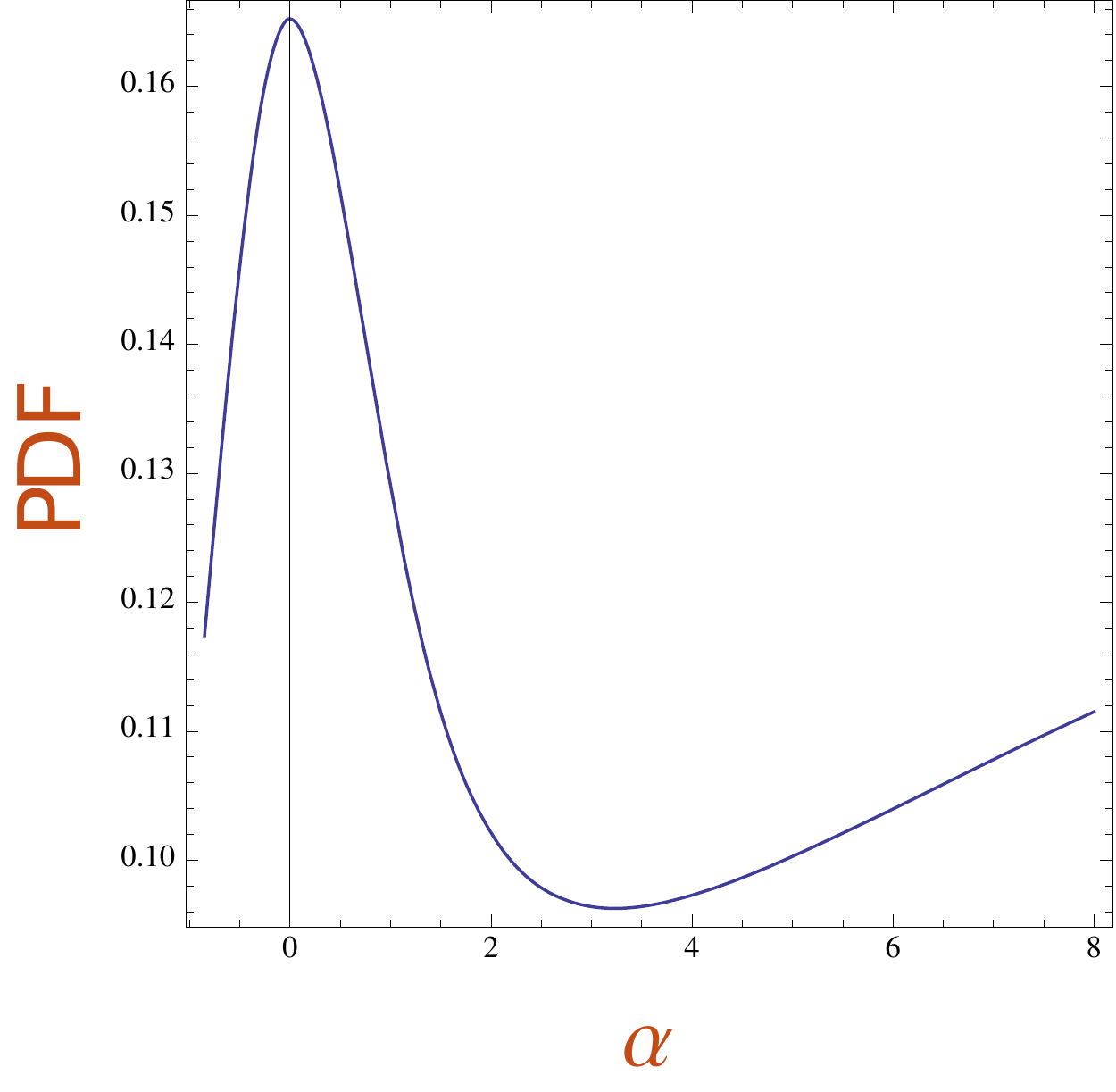}
\hfill
\includegraphics[width=.225\textwidth]{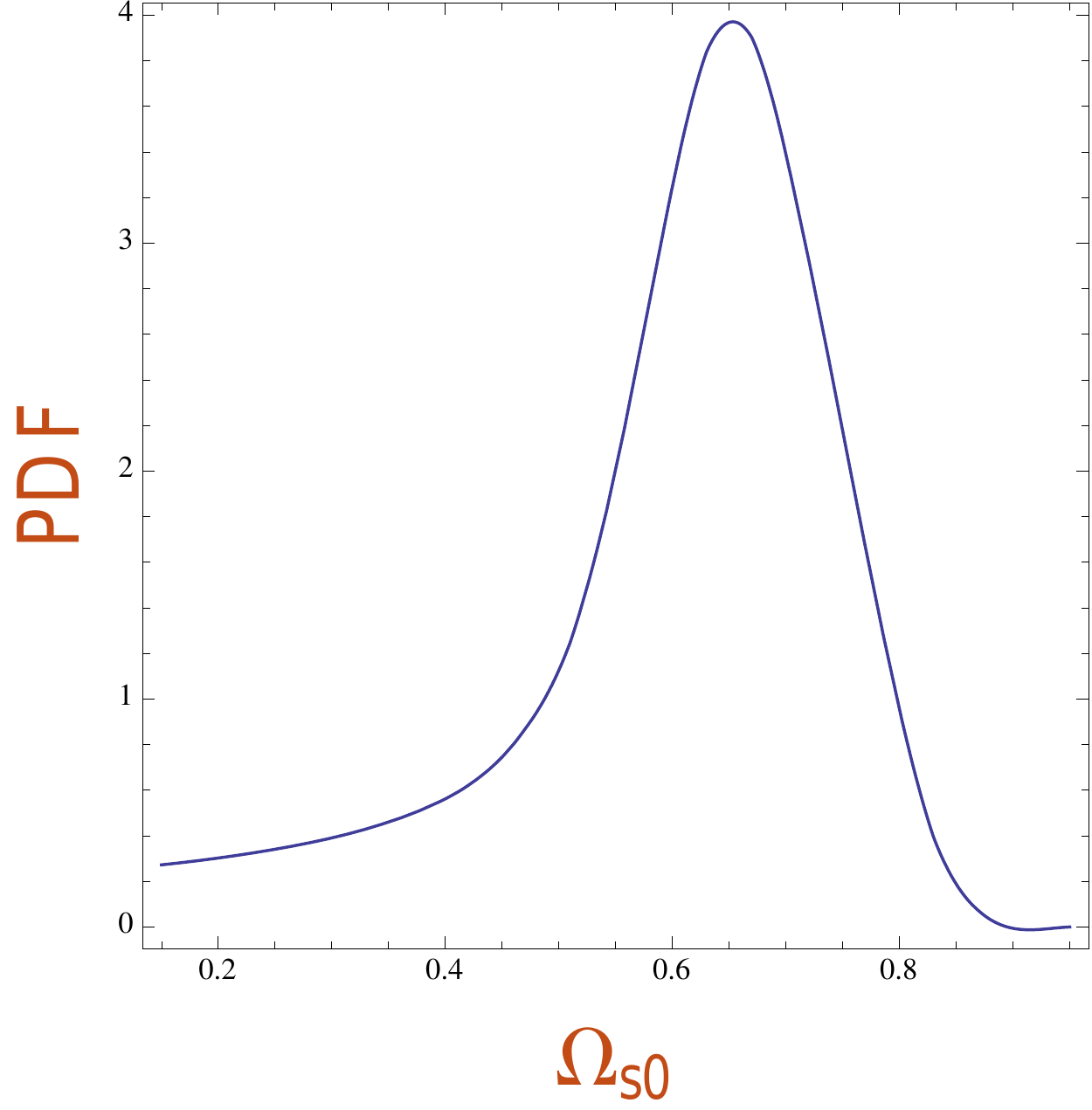}
\hfill
\includegraphics[width=.225\textwidth]{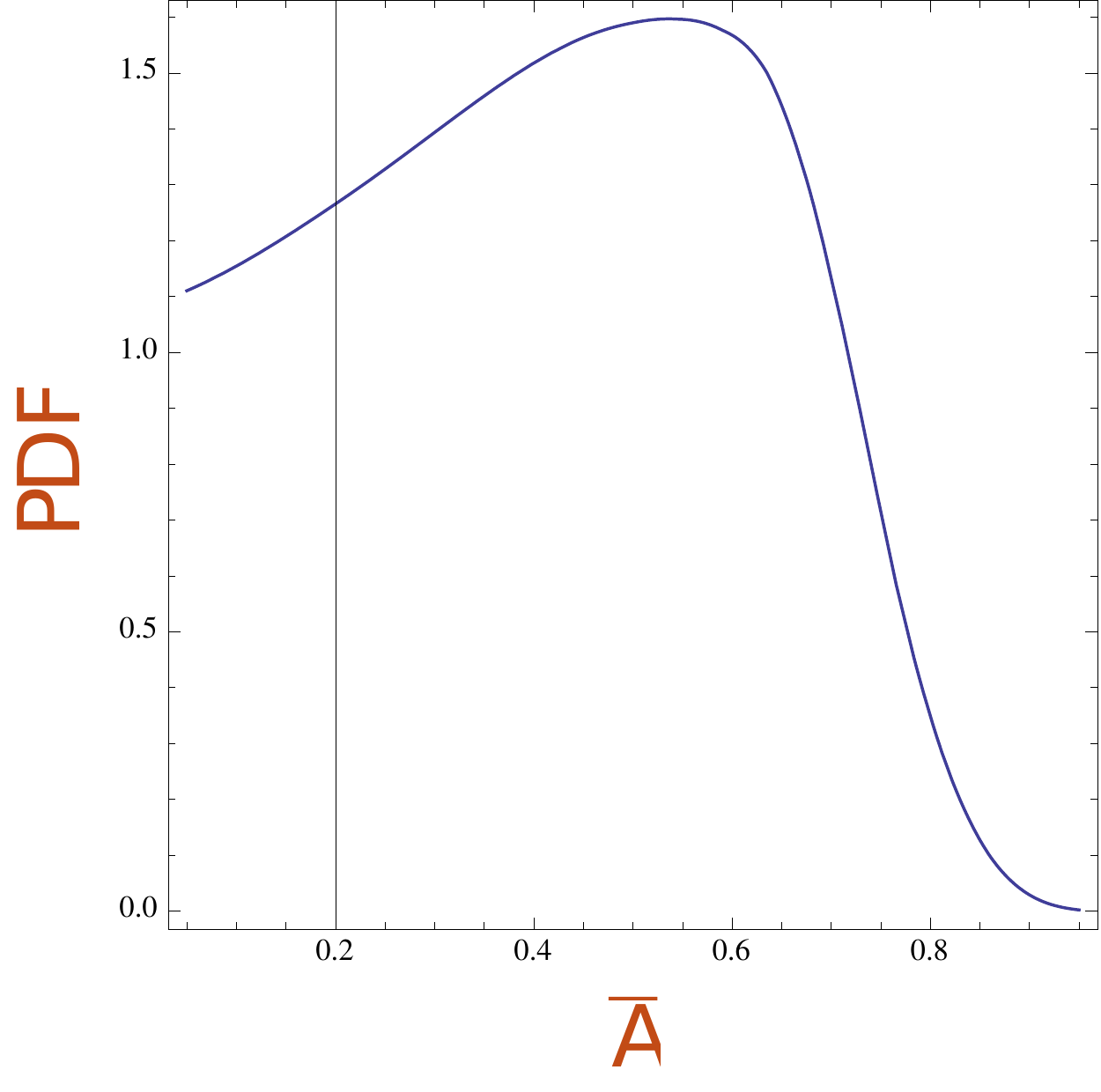}
\hfill
\includegraphics[width=.225\textwidth]{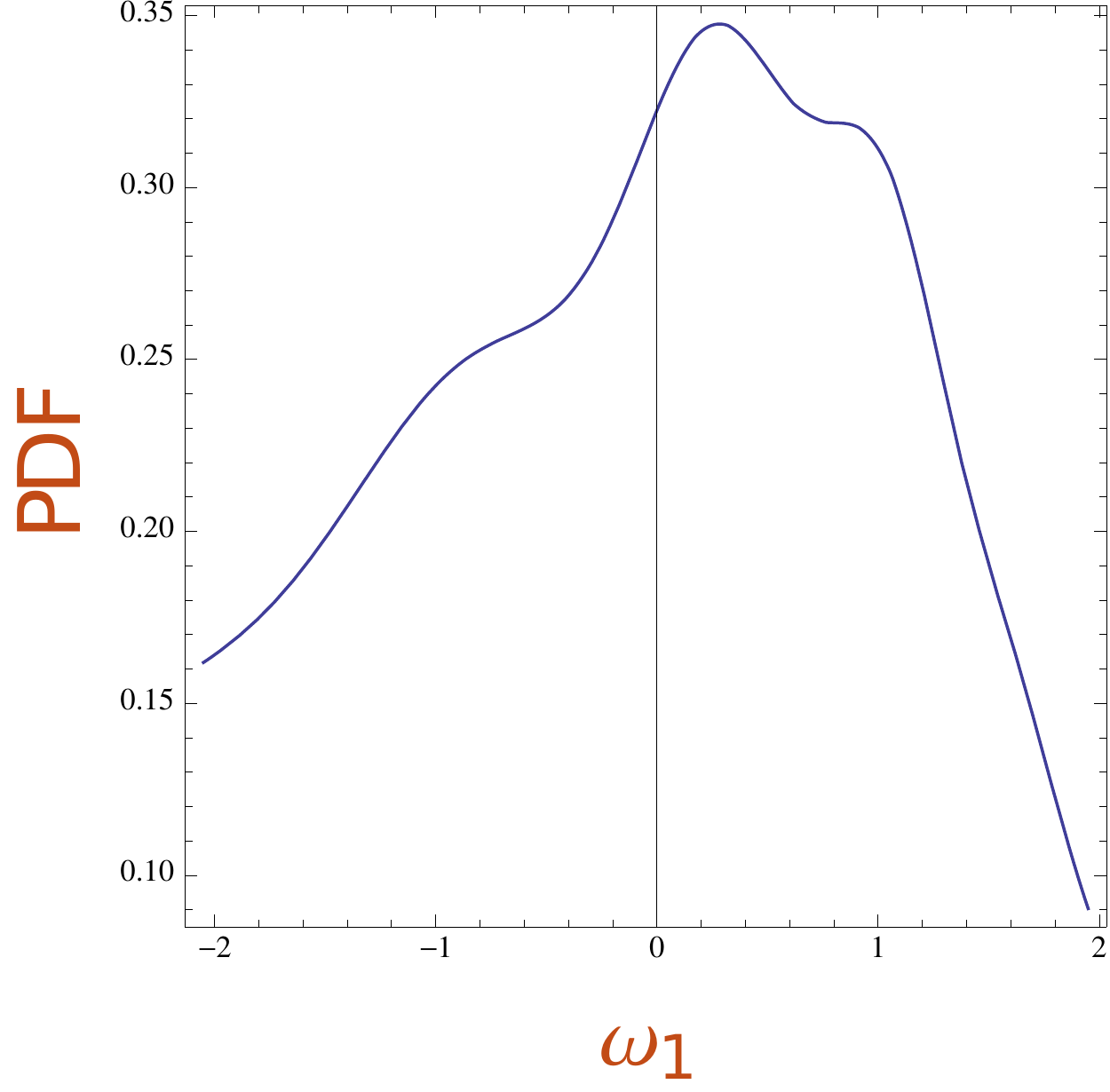}
\caption{One-dimensional PDFs for the 5-parameter CGSF model with $\omega_0 =-1$ and $h$ fixed to its best-fit value.}
\end{figure}

\begin{figure}
\label{2}
\centering
\includegraphics[width=.225\textwidth]{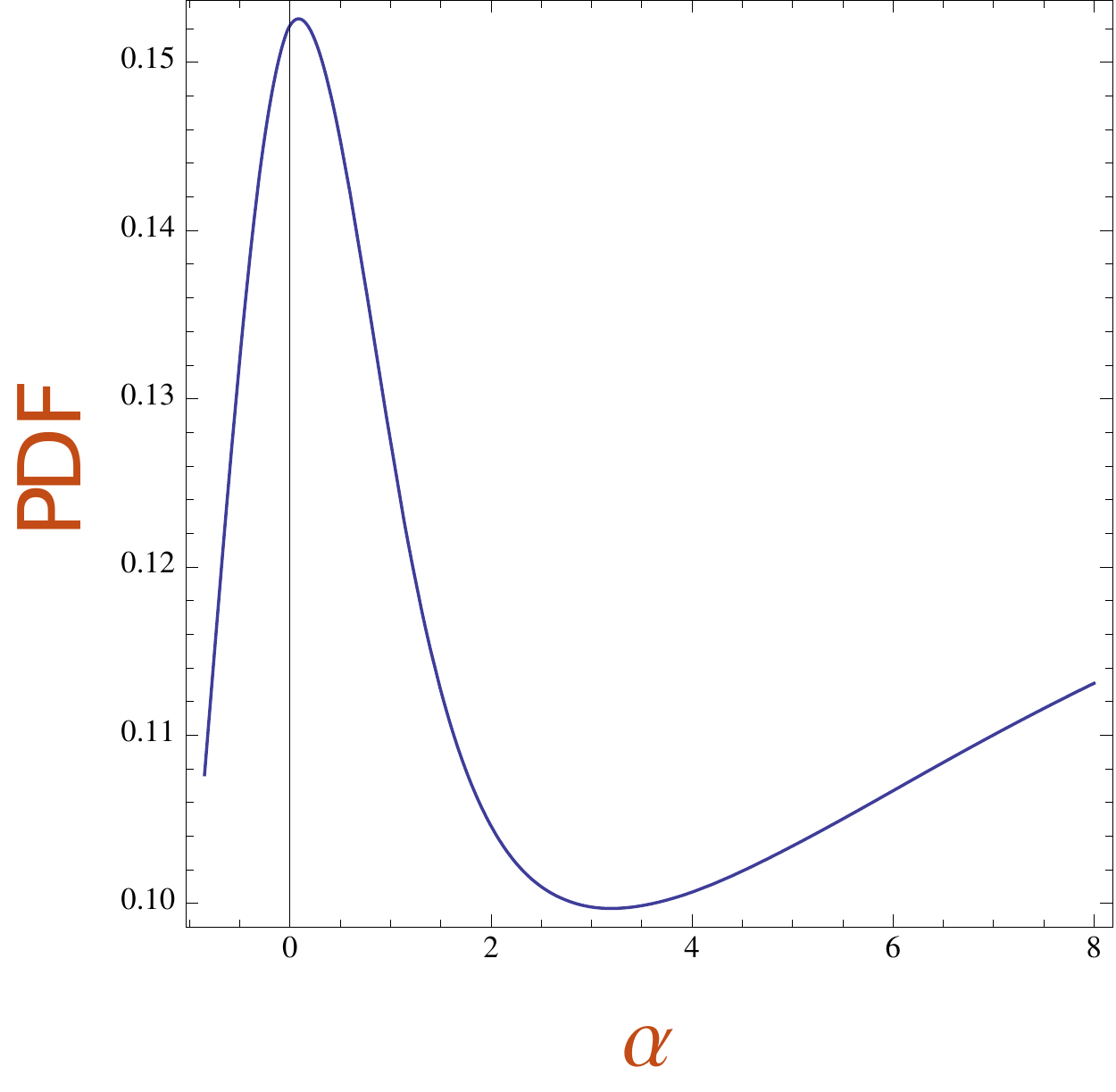}
\hfill
\includegraphics[width=.225\textwidth]{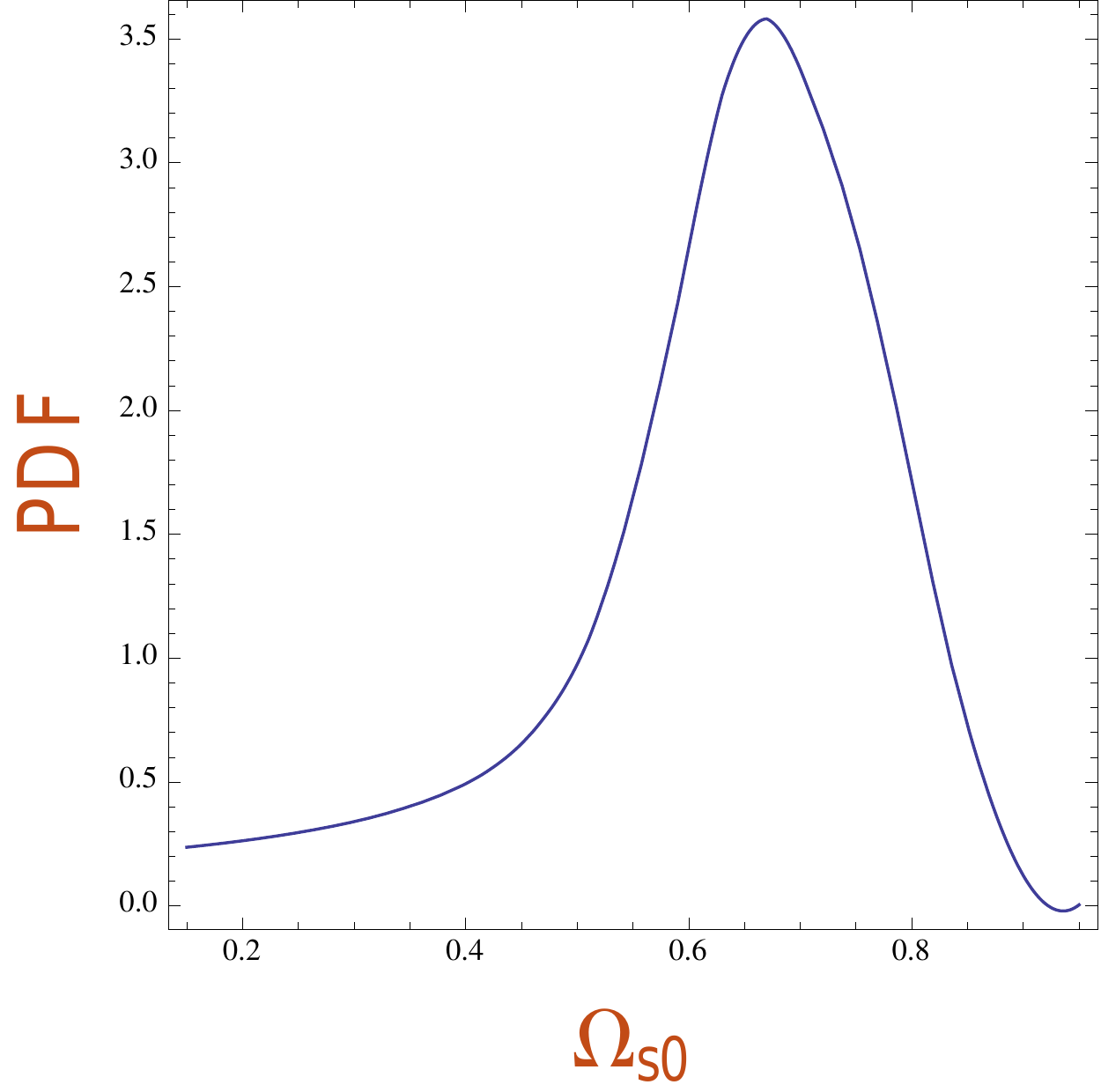}
\hfill
\includegraphics[width=.225\textwidth]{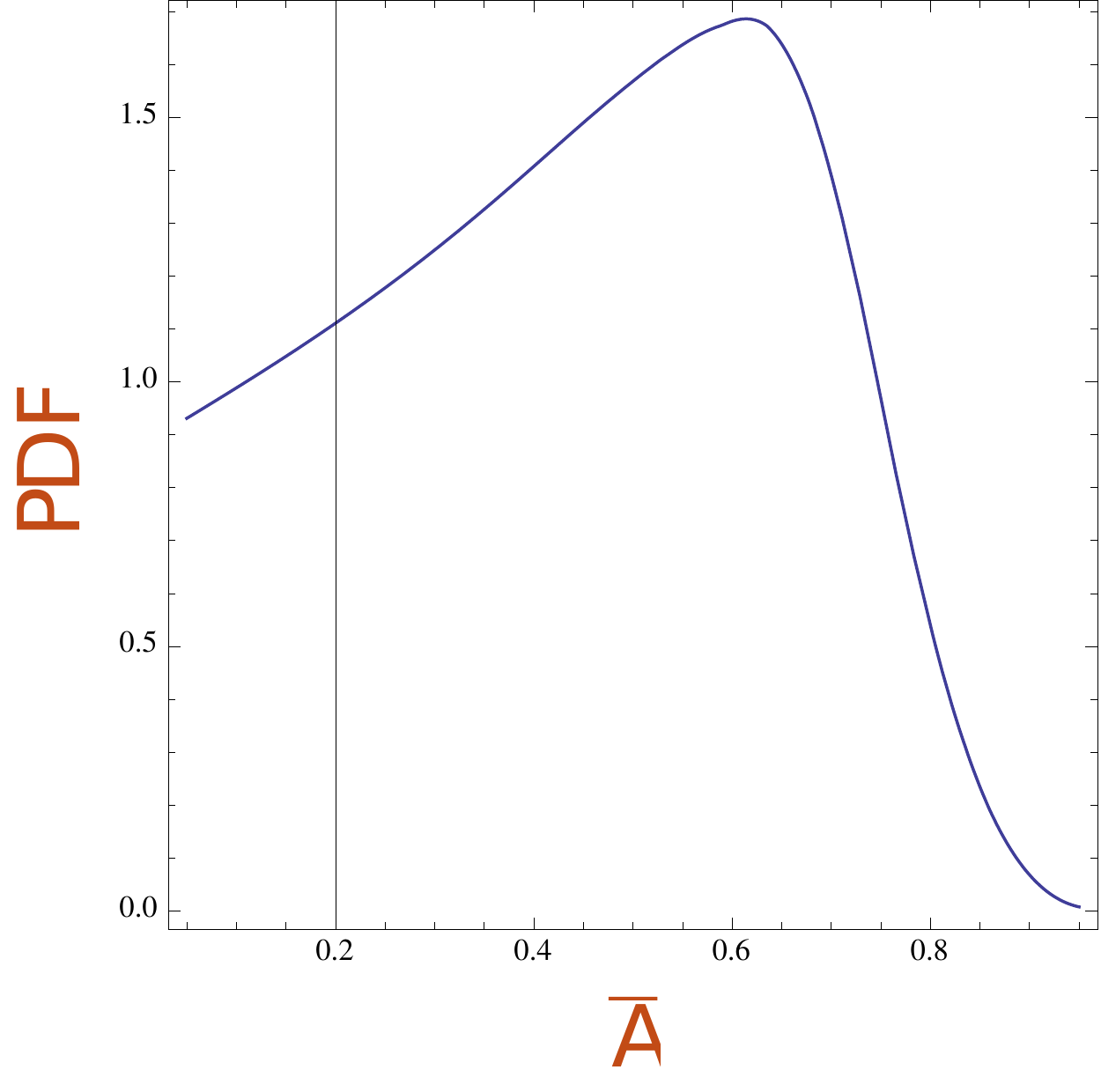}
\hfill
\includegraphics[width=.225\textwidth]{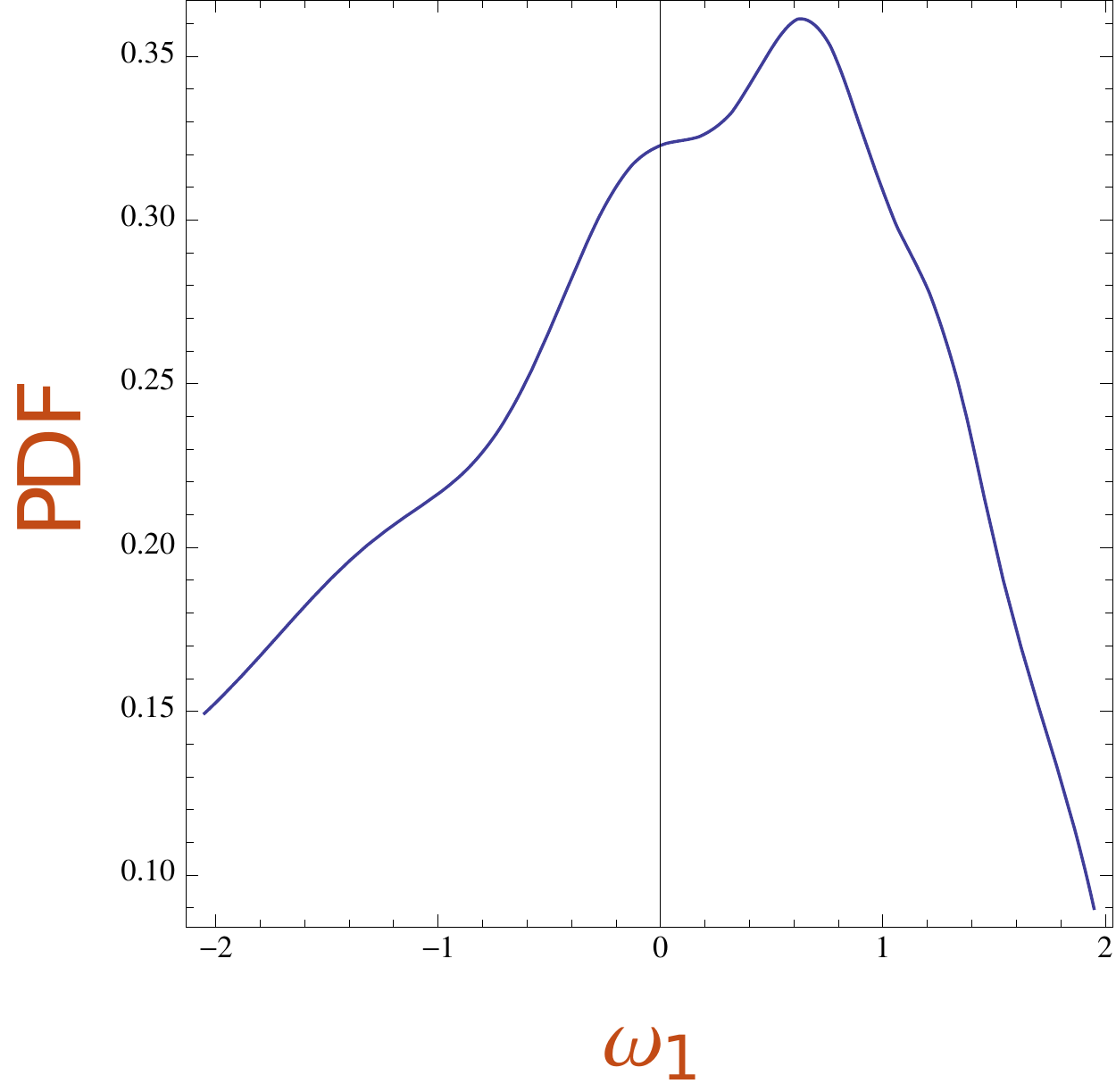}
\caption{One-dimensional PDFs for the 5-parameter CGSF model with $\omega_0 =-0.95$ and $h$ fixed to its best-fit  value.}
\end{figure}

\begin{figure}
\label{3}
\centering
\includegraphics[width=.225\textwidth]{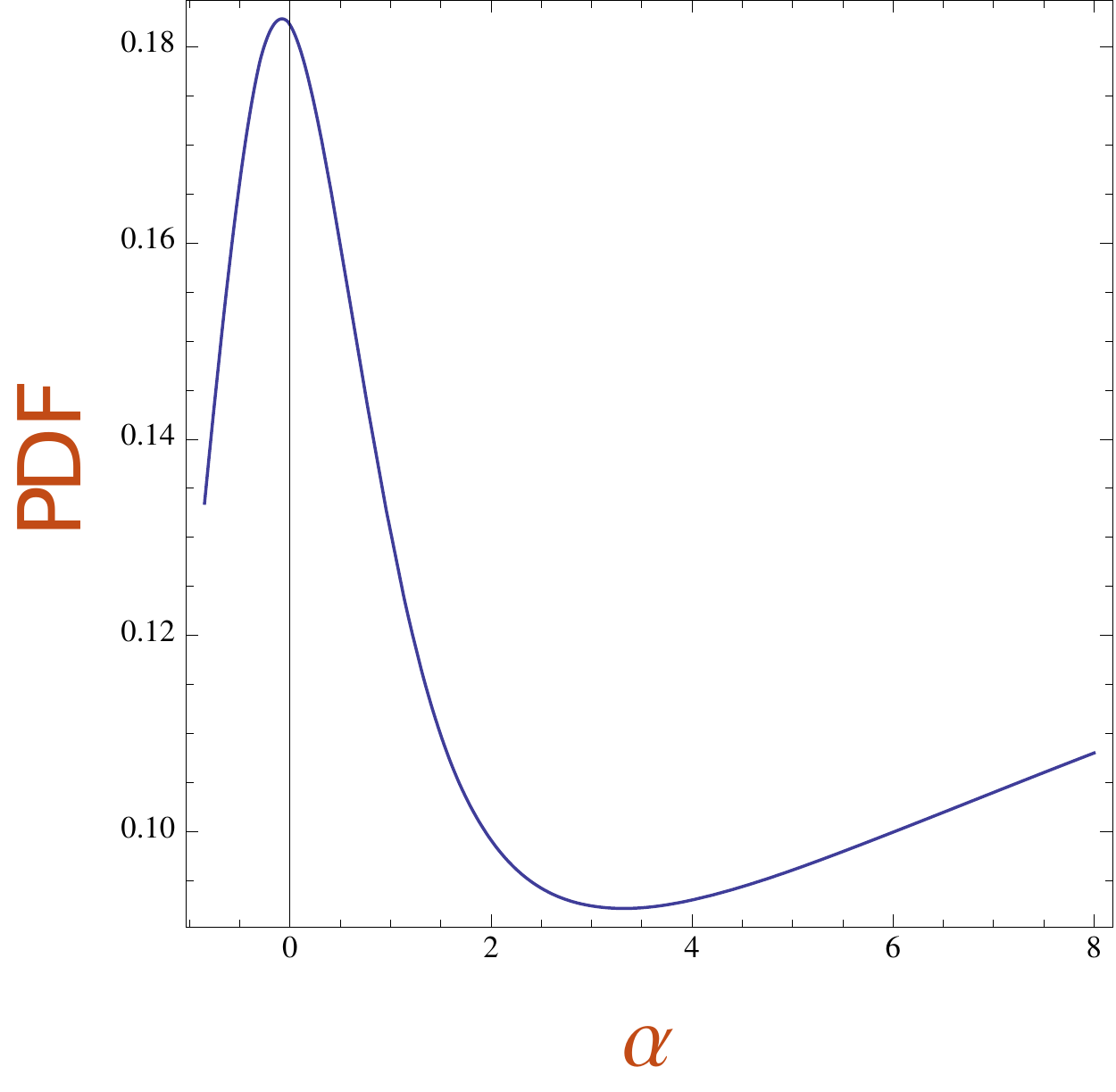}
\hfill
\includegraphics[width=.225\textwidth]{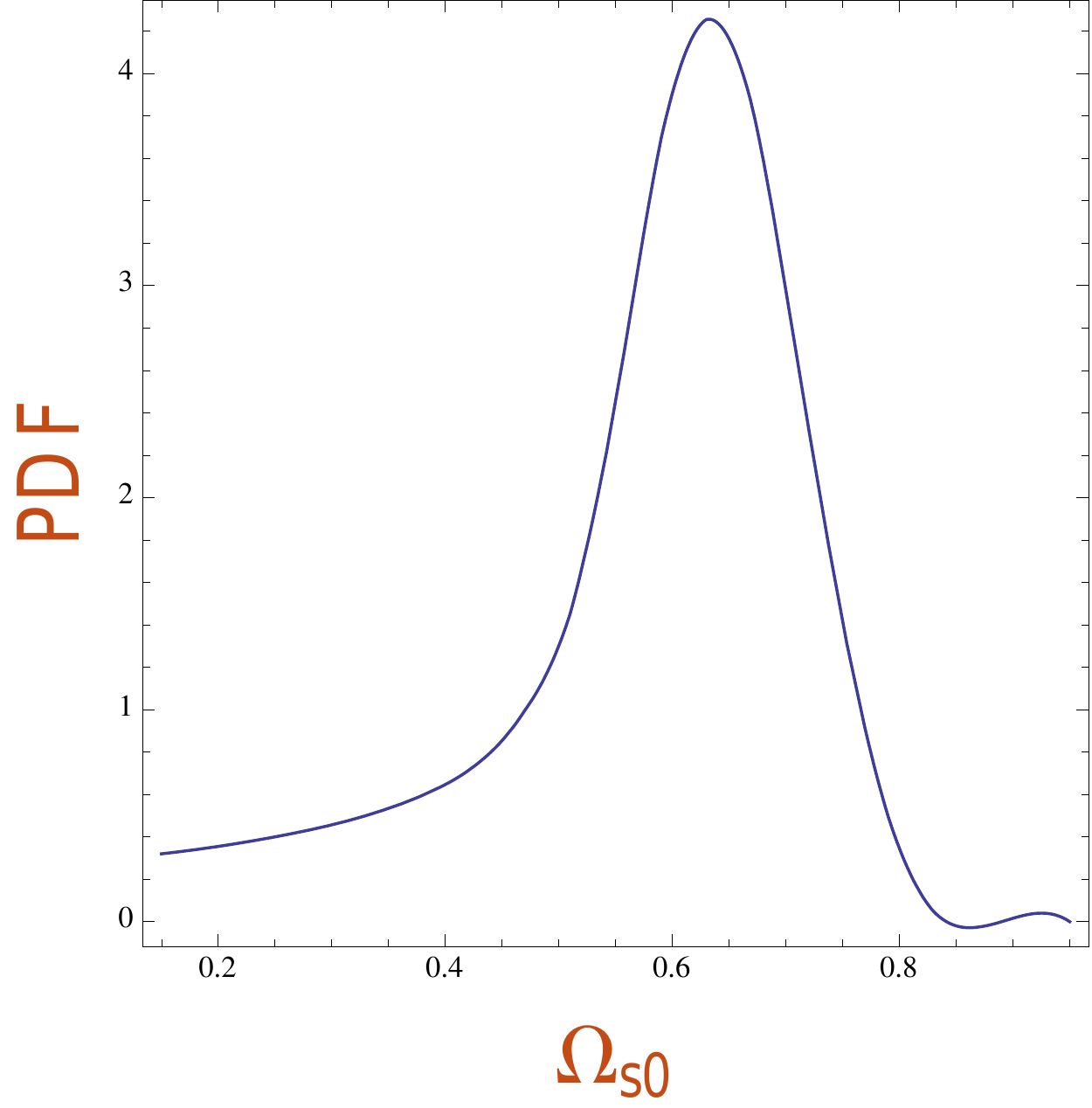}
\hfill
\includegraphics[width=.225\textwidth]{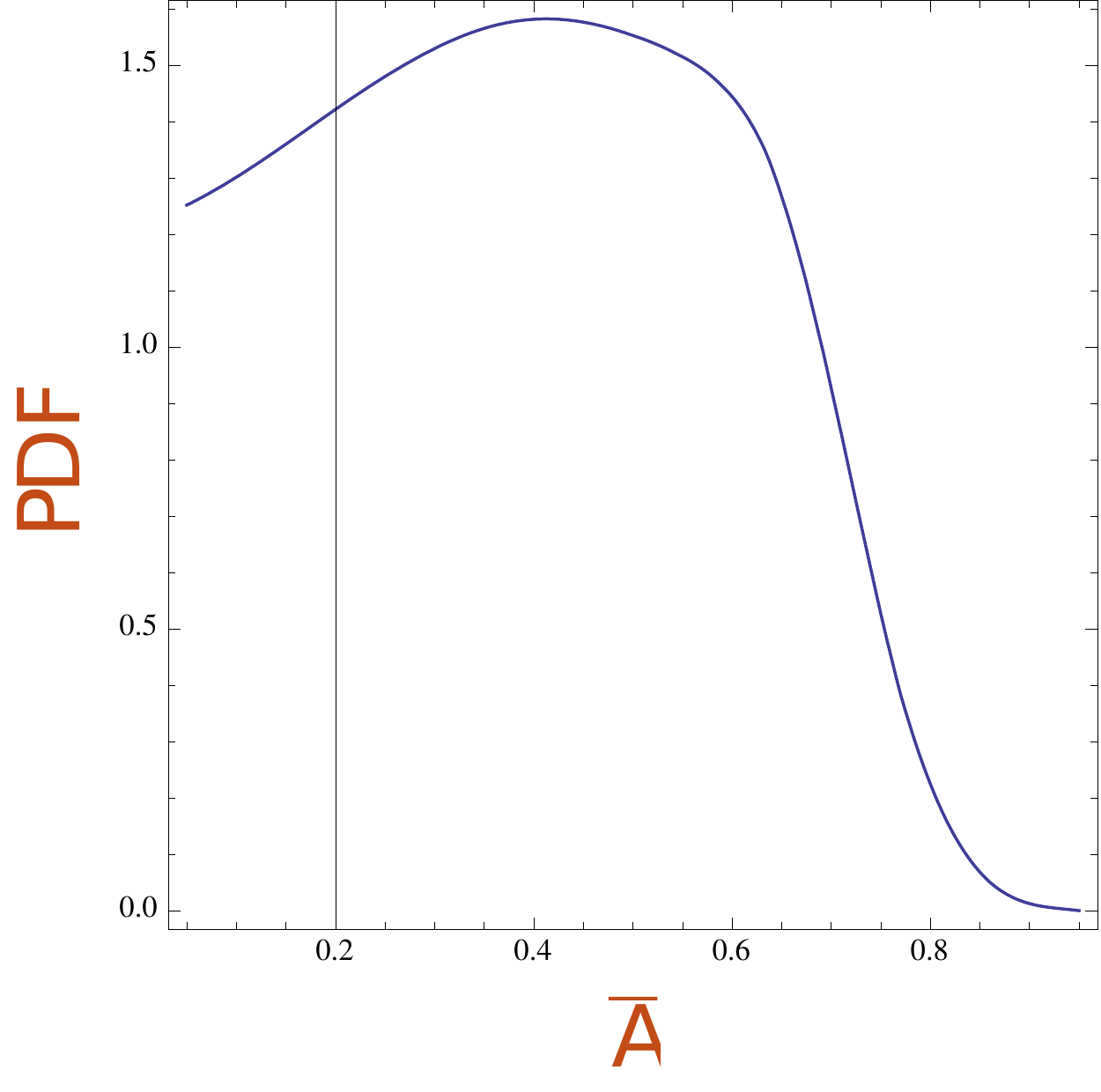}
\hfill
\includegraphics[width=.225\textwidth]{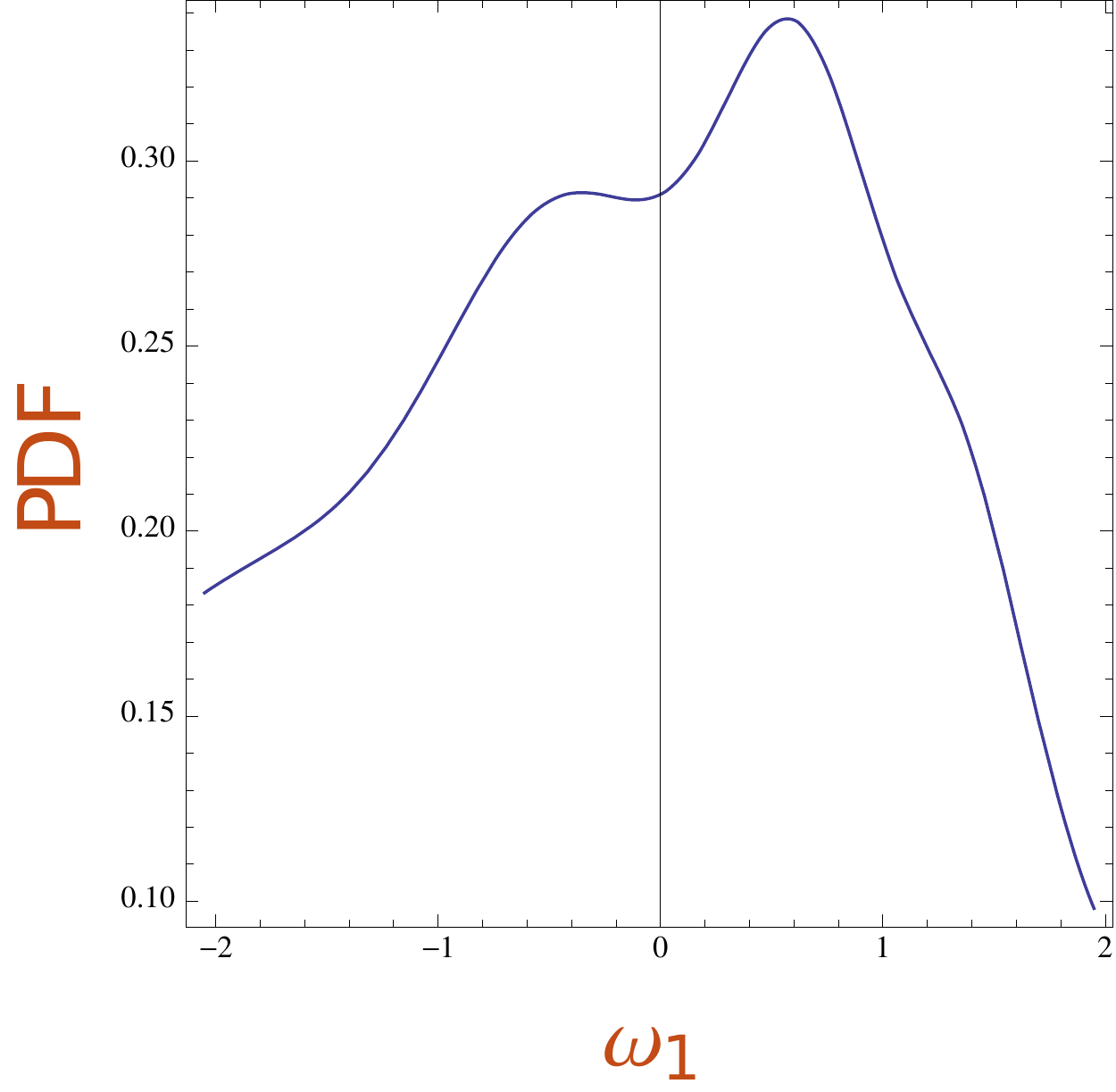}
\caption{One-dimensional PDFs for the 5-parameter CGSF model with $\omega_0 =-1.05$ and $h$ fixed to its best-fit value.}
\end{figure}

\begin{figure}
\label{chap}
\centering
\includegraphics[width=.225\textwidth]{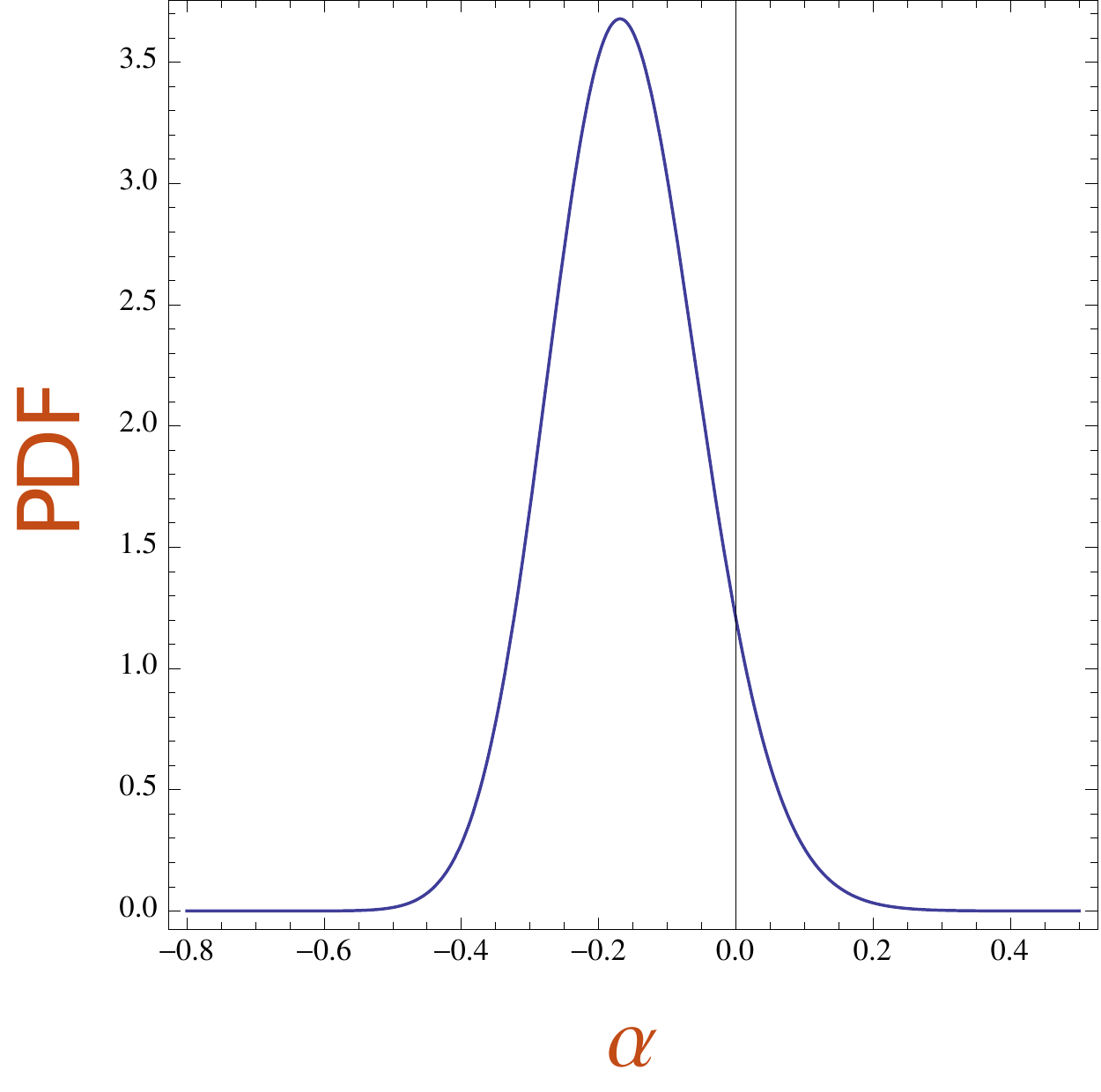}
\hfill
\includegraphics[width=.225\textwidth]{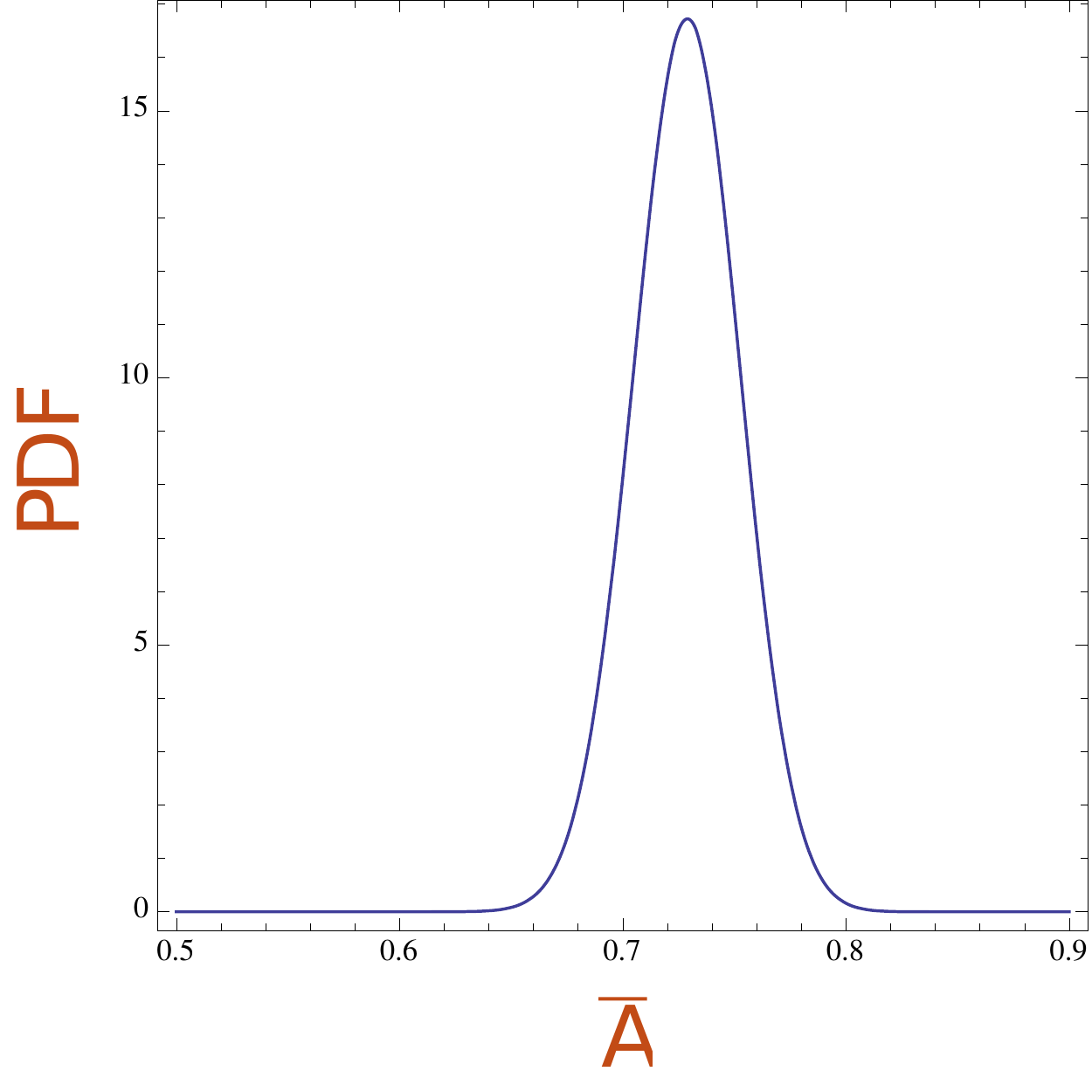}
\hfill
\includegraphics[width=.225\textwidth]{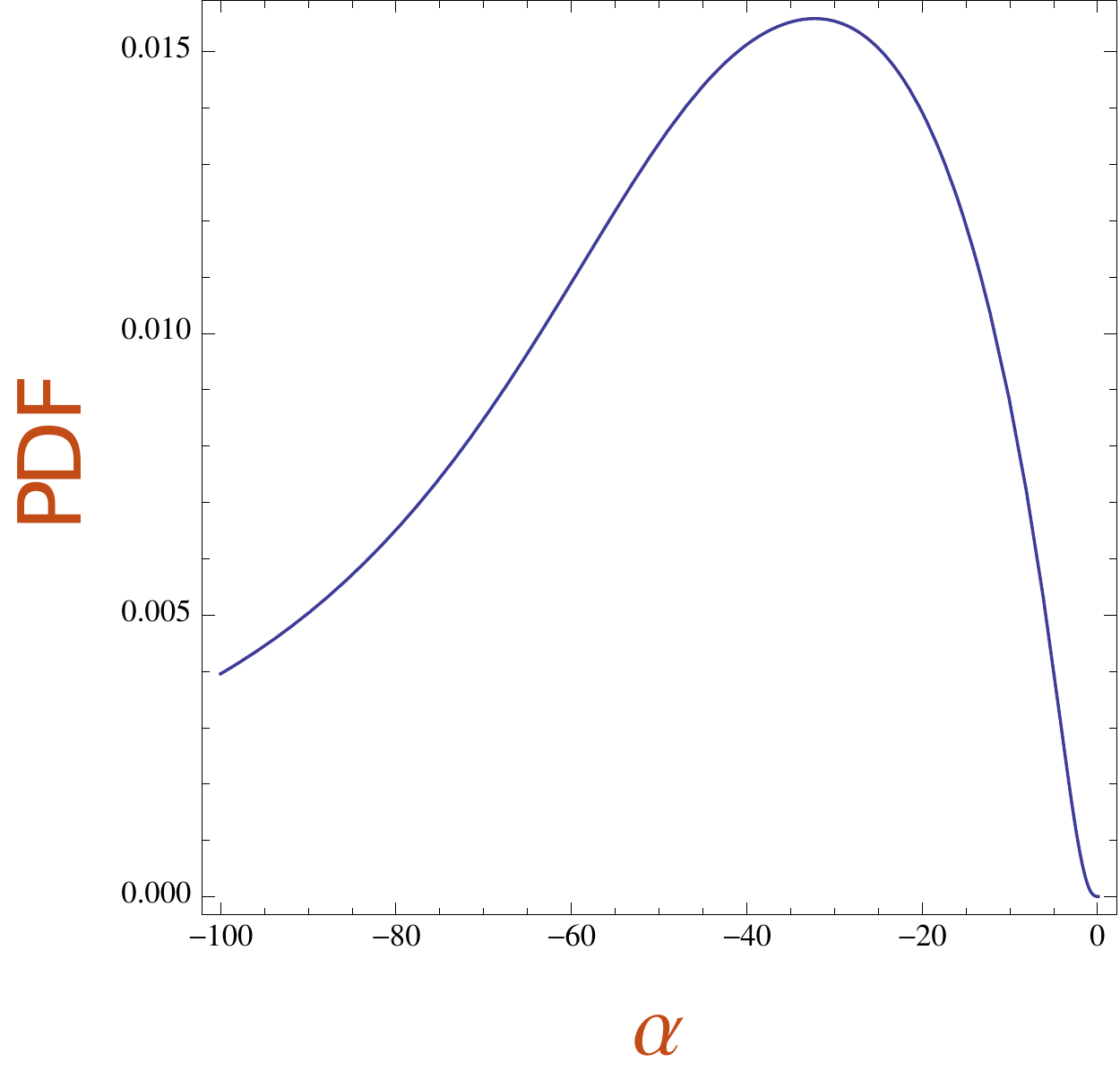}
\hfill
\includegraphics[width=.225\textwidth]{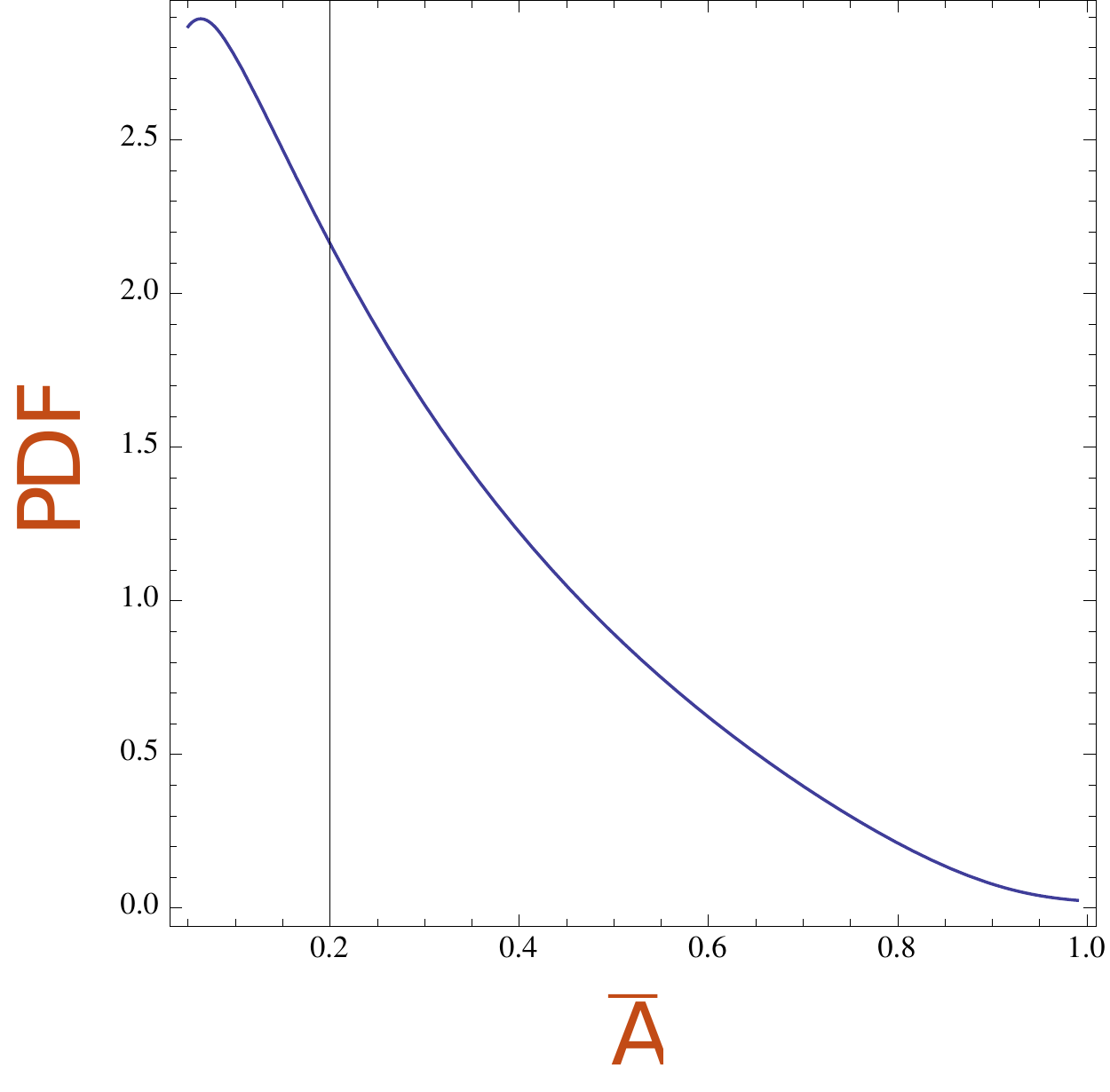}
\caption{PDFs for the pure GCG (no SF) with fixed baryon fraction $\Omega_{m0} = 0.04$ and $h$  value (left two figures)
and with
an additional fixed  matter fraction $\Omega_{m0} = 0.3$ (right figures).}
\end{figure}

\begin{figure}
\label{quint}
\centering
\includegraphics[width=.3\textwidth]{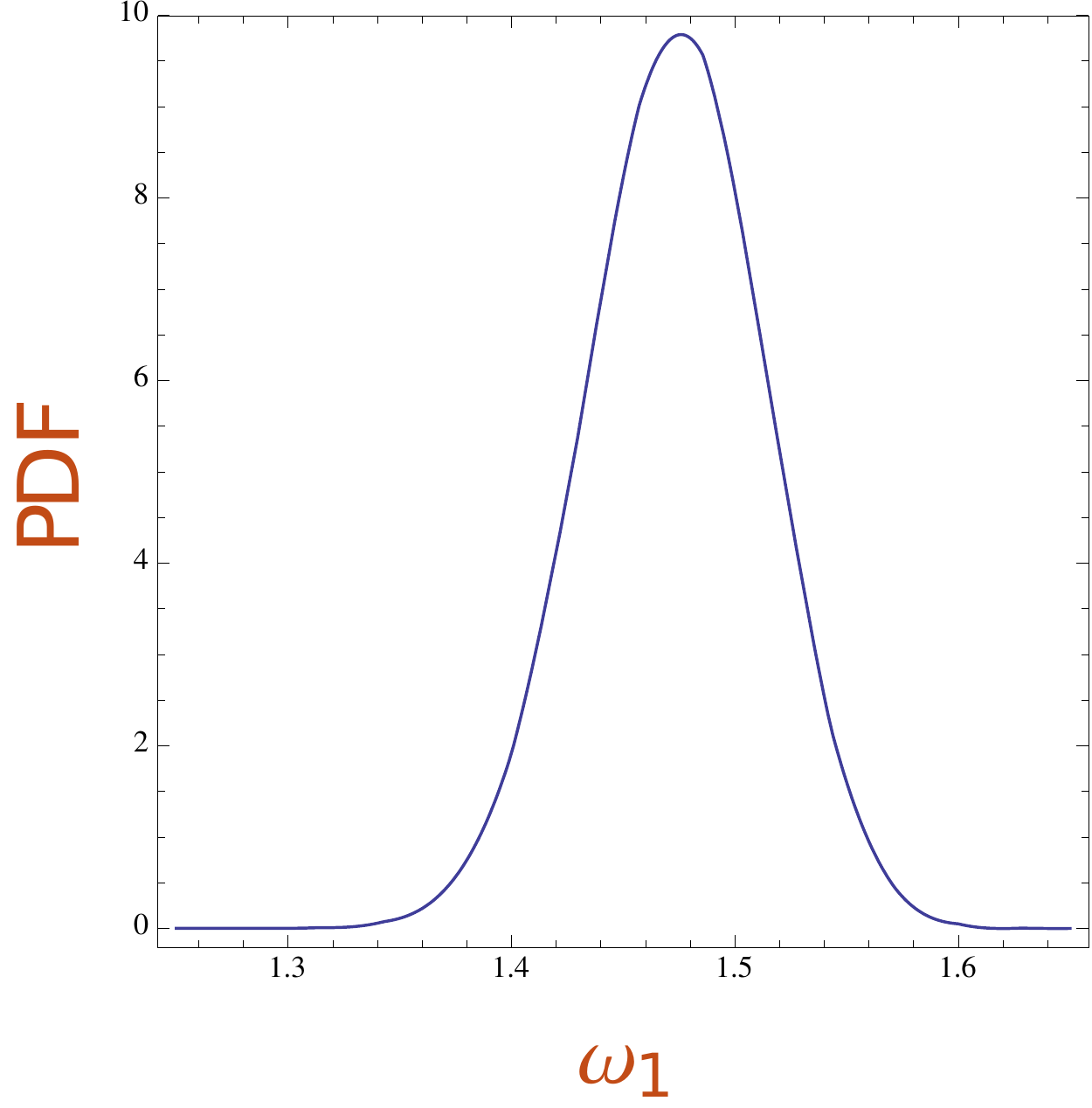}
\hfill
\includegraphics[width=.3\textwidth]{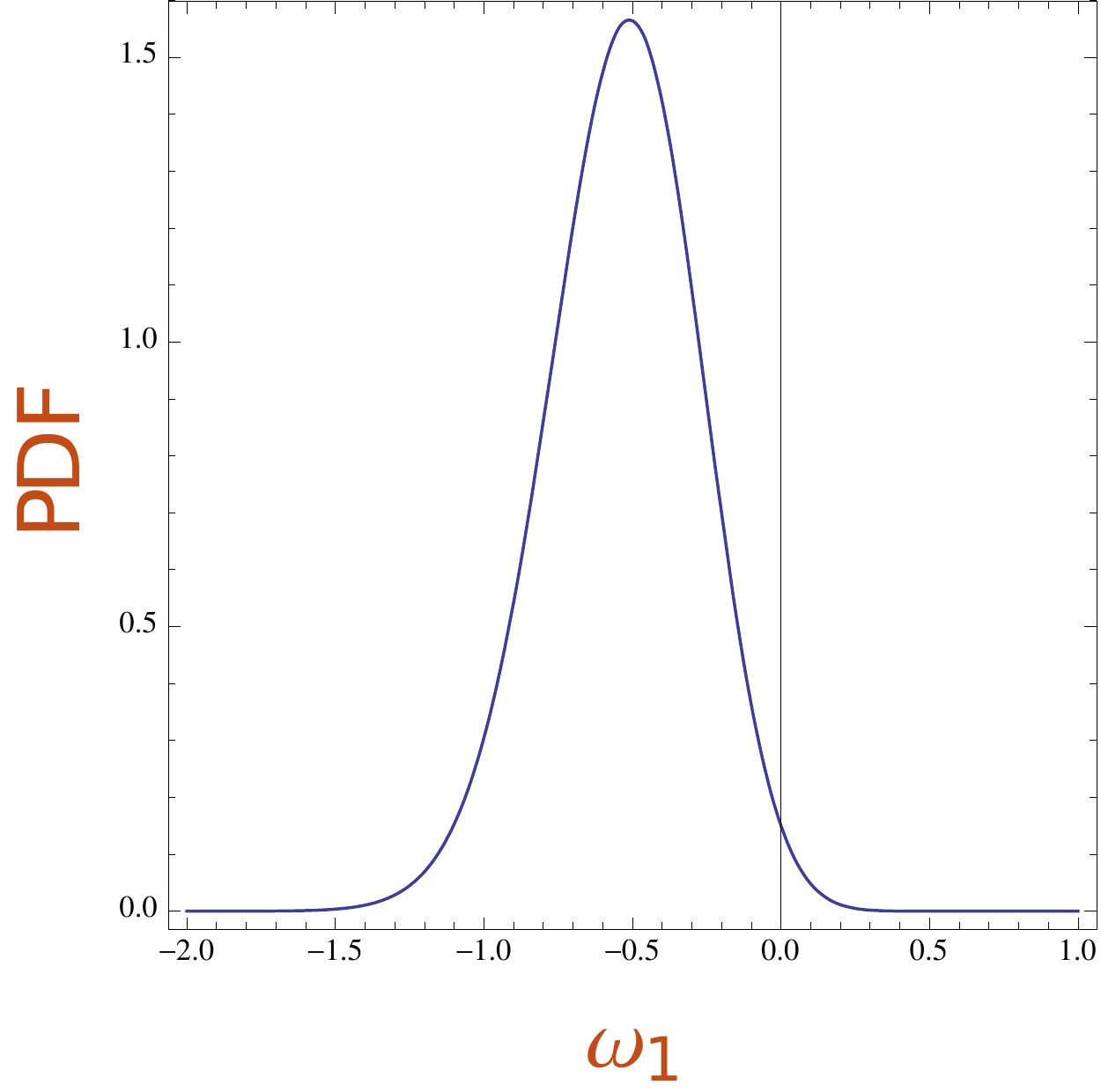}
\caption{PDFs for the pure SF (no GCG) for $\omega_0 =-1$ with fixed baryon fraction $\Omega_{m0} = 0.04$ (equivalent to $\Omega_{s0} = 0.96$), marginalized over $h$(left figure) and with
an additional fixed  matter fraction $\Omega_{m0} = 0.30$ (equivalent to $\Omega_{s0} = 0.70$) (right figure).
}
\end{figure}

\section{Matter perturbations}
\label{perturbations}


In the absence of anisotropic stresses and in the longitudinal gauge scalar metric perturbations
are described by the line element
\begin{equation}
\mbox{d}s^{2} = - \left(1 + 2 \phi\right)\mbox{d}t^2  +
a^2\left(1-2\phi\right)\delta _{\alpha \beta} \mbox{d}x^\alpha\mbox{d}x^\beta.\label{ds}
\end{equation}
Denoting first-order variables by a hat symbol, the perturbed
time components of the four-velocities are
\begin{equation}
\hat{u}_{0} = \hat{u}^{0} = \hat{u}_{A}^{0}  = \frac{1}{2}\hat{g}_{00} = - \phi\ .
\label{u0}
\end{equation}
Under this condition the total first-order energy perturbation  is the sum of the first-order perturbations of the components of the cosmic medium,
\begin{equation}\label{}
\hat{\rho} = \hat{\rho}_{m} + \hat{\rho}_{c} + \hat{\rho}_{s}.
\end{equation}
Introducing the fractional quantities
\begin{equation}\label{}
\delta = \frac{\hat{\rho}}{\rho}, \quad\delta_{m} = \frac{\hat{\rho}_{m}}{\rho_{m}}, \quad\delta_{c} = \frac{\hat{\rho}_{c}}{\rho_{c}}, \quad\delta_{s} = \frac{\hat{\rho}_{s}}{\rho_{s}},
\end{equation}
one has
\begin{equation}\label{}
\delta = \Omega_{m}\delta_{m} + \Omega_{c}\delta_{c} + \Omega_{s}\delta_{s},
\end{equation}
where
\begin{eqnarray}\label{}
 \Omega_{m} &=& \Omega_{m0}\frac{H_{0}^{2}}{H^{2}}a^{-3},\\
\Omega_{c} &=& \Omega_{c0}\frac{H_{0}^{2}}{H^{2}}\left[\bar{A} + \left(1 - \bar{A}\right)a^{-3\left(1+ \alpha\right)}\right]^{\frac{1}{1+\alpha}},\\
\Omega_{s} &=& \Omega_{s0}\frac{H_{0}^{2}}{H^{2}}a^{-3\left(1+\omega_{0}+w_{1}\right)}
e^{3\omega_{1}\left(a-1\right)}
\end{eqnarray}
with $\frac{H_{0}^{2}}{H^{2}}$ from (\ref{hubble}).

The equation for the fractional matter perturbations $\delta_{m}$ in the  quasi-static,
sub-horizon approximation is
\begin{equation}\label{dddeltam}
\ddot{\delta}_{m} + 2H\dot{\delta}_{m} + \frac{k^{2}}{a^2}\phi= 0,
\end{equation}
where $k$ is the comoving wave number. Einstein's field equations relate $\phi$ to the (comoving) energy-density perturbations via the Poisson equation
\begin{equation}\label{Newtonsim}
\frac{k^{2}}{a^{2}}\phi = -4 \pi G\rho\delta .
\end{equation}
Combining (\ref{dddeltam}) and (\ref{Newtonsim}) does not result in a closed equation for the matter perturbations, since $\delta_{m}$ is generally coupled to the perturbations of the other components.
Formally, we may write
\begin{equation}\label{}
\delta = \left[\Omega_{m} + \Omega_{c}\frac{\delta_{c}}{\delta_{m}} + \Omega_{s}\frac{\delta_{s}}{\delta_{m}}\right]\delta_{m}.
\end{equation}
Only for known ratios $\frac{\delta_{c}}{\delta_{m}}$  and $\frac{\delta_{s}}{\delta_{m}}$ there would be a a closed second-order equation for $\delta_{m}$.
This means, to obtain $\delta_{m}$ one has to solve the entire coupled dynamics of $\delta_{m}$, $\delta_{c}$ and $\delta_{s}$.
To get a rough idea of how the fluctuations of the GCG and the SF components affect the matter fluctuations we introduce the simple  parametrizations
\begin{equation}\label{para}
\frac{\delta_{c}}{\delta_{m}} = \mu a^{c},\qquad \frac{\delta_{s}}{\delta_{m}} = \nu a^{s},
\end{equation}
in which the constant parameters $\mu$ and $\nu$ represent the values of $\frac{\delta_{c}}{\delta_{m}}$  and $\frac{\delta_{s}}{\delta_{m}}$, respectively, at the present time. The powers $c$ and $s$ account for deviations in the dynamics of the perturbations in terms of their dependence on the scale factor.
While such parametrization does not replace the necessity of an exact solution, we expect it to provide us with a provisional insight concerning the complicated coupled perturbation dynamics, at least close to $a=1$,  i.e., at small redshift.
Under such condition the equation for $\delta_{m}$ becomes
\begin{equation}\label{dddeltamfin}
\ddot{\delta}_{m} + 2H\dot{\delta}_{m} -4 \pi G\rho\left[\Omega_{m} + \mu a^{c}\Omega_{c} + \nu a^{s} \Omega_{s}\right]\delta_{m}= 0,
\end{equation}
equivalent to
\begin{equation}\label{dprpr}
\delta_{m}^{\prime\prime} + \frac{3}{2}\left(1-\omega\right)\frac{\delta_{m}^{\prime}}{a}
-\frac{3}{2}\left[\Omega_{m} + \mu a^{c}\Omega_{c} + \nu a^{s} \Omega_{s}\right]\frac{\delta_{m}}{a^{2}} =0,
\end{equation}
with
\begin{equation}\label{}
\omega = \frac{p}{\rho} = \omega_{c}\Omega_{c} + \omega_{s}\Omega_{s},
\end{equation}
where $\omega_{C}$ is defined in (\ref{wC}) and $\omega_{S} = \omega_{0} + \omega_{1}\left(1-a\right)$.
All background coefficients in  equation (\ref{dprpr}) are analytically known. This allows us to study the behavior of matter perturbations for different combinations of the parameters $\mu$ and $\nu$  as well as of the powers $c$ and $s$ for the best-fit values of the background quantities.

In Figs.~6 - 9 we visualize the scale-factor dependence of the fractional matter perturbation $\delta_{m}$ for various combinations of the parameters introduced in (\ref{para}). Fig.~6 shows how variations  of the parameter $\mu$ influence the matter growth. For comparison we include also the curve for the $\Lambda$CDM model.
Apparently, $\mu$ values of the order of one make the matter perturbations similar to those of the standard model.
For $\mu \lesssim 1$ there is a clear tendency to a lower growth than predicted by the standard model, i.e., to less structure formation.
A value of $\mu \approx 1$ means that currently the fluctuations in the GCG component are of the same order as the baryonic matter fluctuations. Although there is no separate CDM component in the present configuration, the GCG accounts for that part of the dynamics which in the standard model is described by CDM. In this sense, for $\mu \approx 1$,
CDM fluctuations as part of the GCG fluctuations are roughly of the same order as the baryonic matter fluctuations. 
Fig.~7 shows the consequence of a variation in the power $c$. Any deviation from $c=0$ leads to a substantial deviation from the standard model.
As to be seen from Fig.~8, even a moderate value of $\nu$ of $\nu=0.5$ together with $\mu=0.5$ leads to a strong deviation from the $\Lambda$CDM curve. This indicates that $\nu$ has to be rather small compared with $\mu$, implying  that fluctuations of the scalar-field component are not relevant on sub-horizon scales which is in accord with the well-known result of \cite{Staro98}.
Fig.~9 confirms that values around $\mu \approx 1$ with $\nu \ll 1$ remain in the vicinity of the $\Lambda$CDM model. While the background is dynamically dominated by the scalar-field component, the fluctuations of this component are irrelevant compared with the fluctuations of the GCG component.

\begin{figure}[h!]
\label{}
\begin{center}
\includegraphics[width=0.5\textwidth]{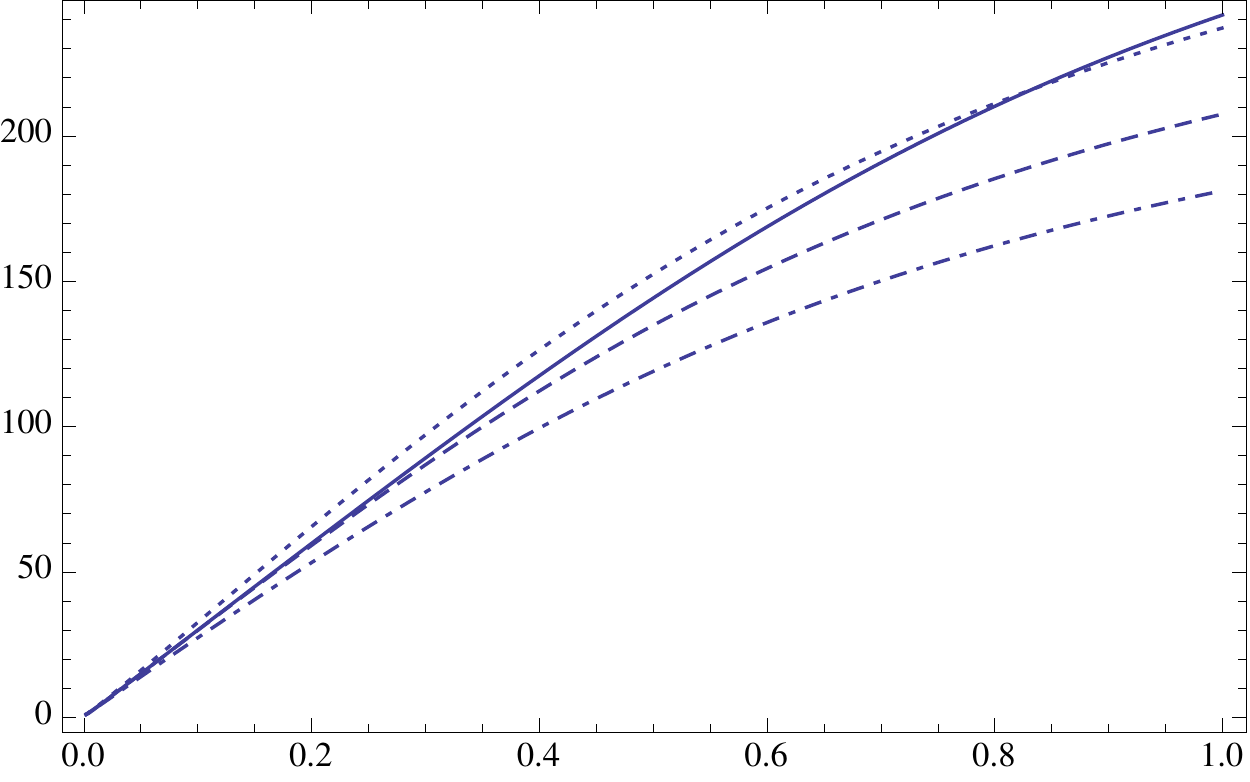}
\label{1}	
\caption{The matter perturbation $\delta_{m}$ as a function of the scale factor $a$ for $\nu = c= s= 0$.
The dotted curve corresponds to $\mu = 1.05$, the dashed curve to $\mu = 1.0$  and the dot-dashed curve to $\mu = 0.95$. The solid curve represents the $\Lambda$CDM model.}
\end{center}
\end{figure}

\begin{figure}[h!]
\label{}
\begin{center}
\includegraphics[width=0.5\textwidth]{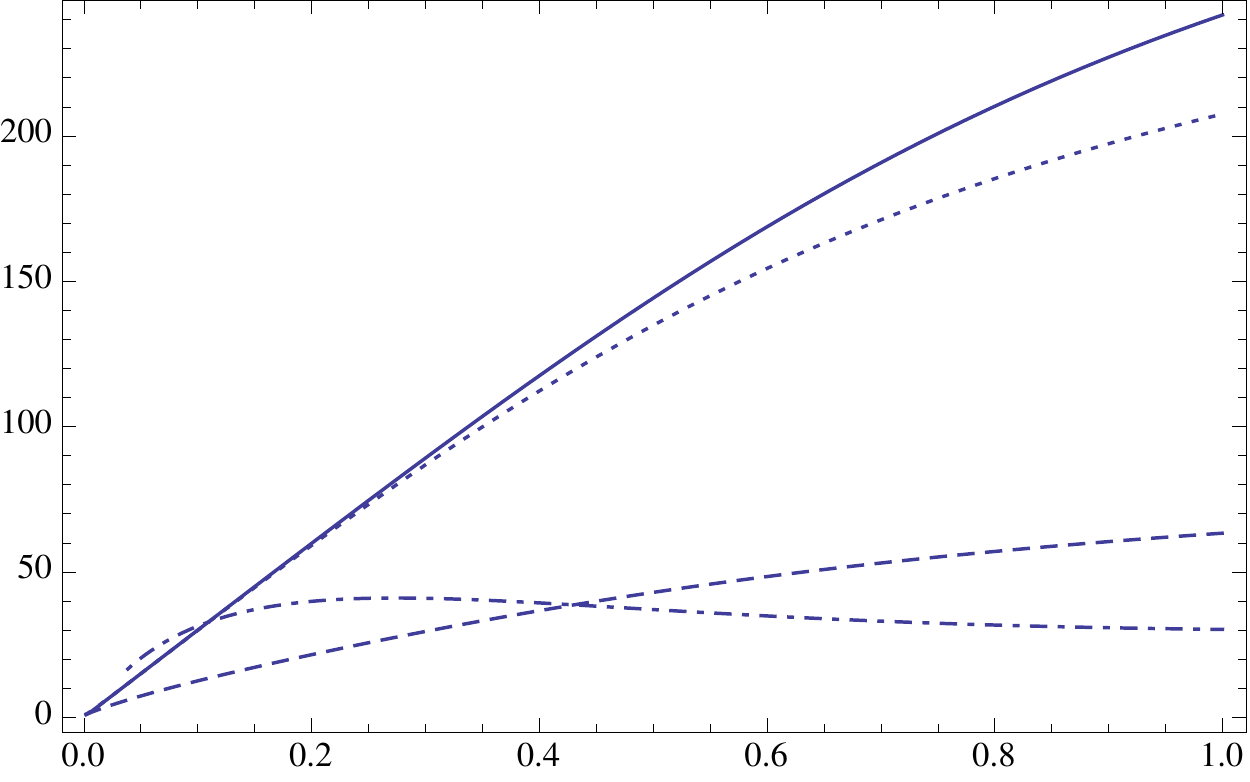}
\label{2}	
\caption{The matter perturbation $\delta_{m}$ as a function of the scale factor $a$  for $\nu = s =0$ and $\mu = 1.0$.
The dotted curve describes $c=0$, the dashed curve $c=0.1$ and the dot-dashed curve $c=-0.1$. Apparently, any non-vanishing $s$ leads to unacceptable strong deviations from the $\Lambda$CDM model (solid curve).}
\end{center}
\end{figure}

\begin{figure}[h!]
\label{}
\begin{center}
\includegraphics[width=0.5\textwidth]{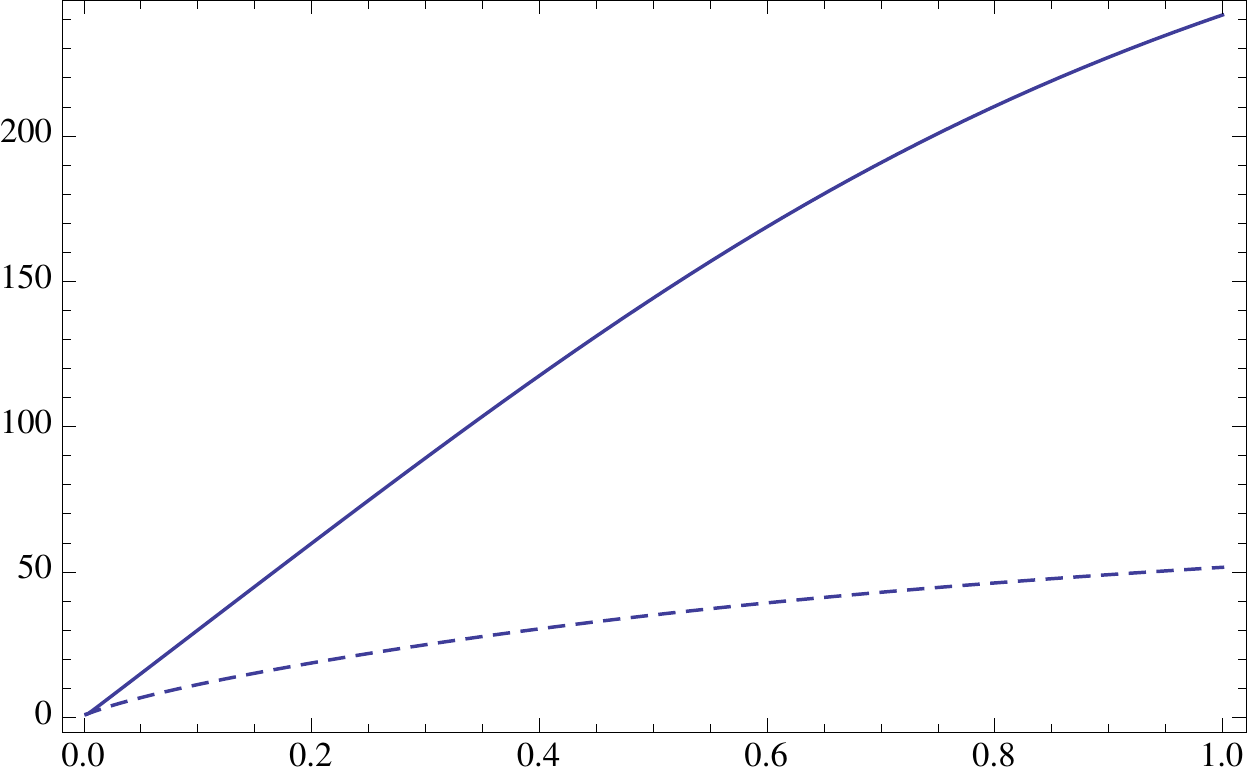}
\label{}	
\caption{The matter perturbation $\delta_{m}$ as a function of the scale factor $a$ for  $\mu = 0.5$ and $\nu = 0.5$. For $\nu$-values of the order of the values for $\mu$ there is substantial disagreement with the $\Lambda$CDM model (solid curve).}
\end{center}
\end{figure}

\begin{figure}[h!]
\label{}
\begin{center}
\includegraphics[width=0.5\textwidth]{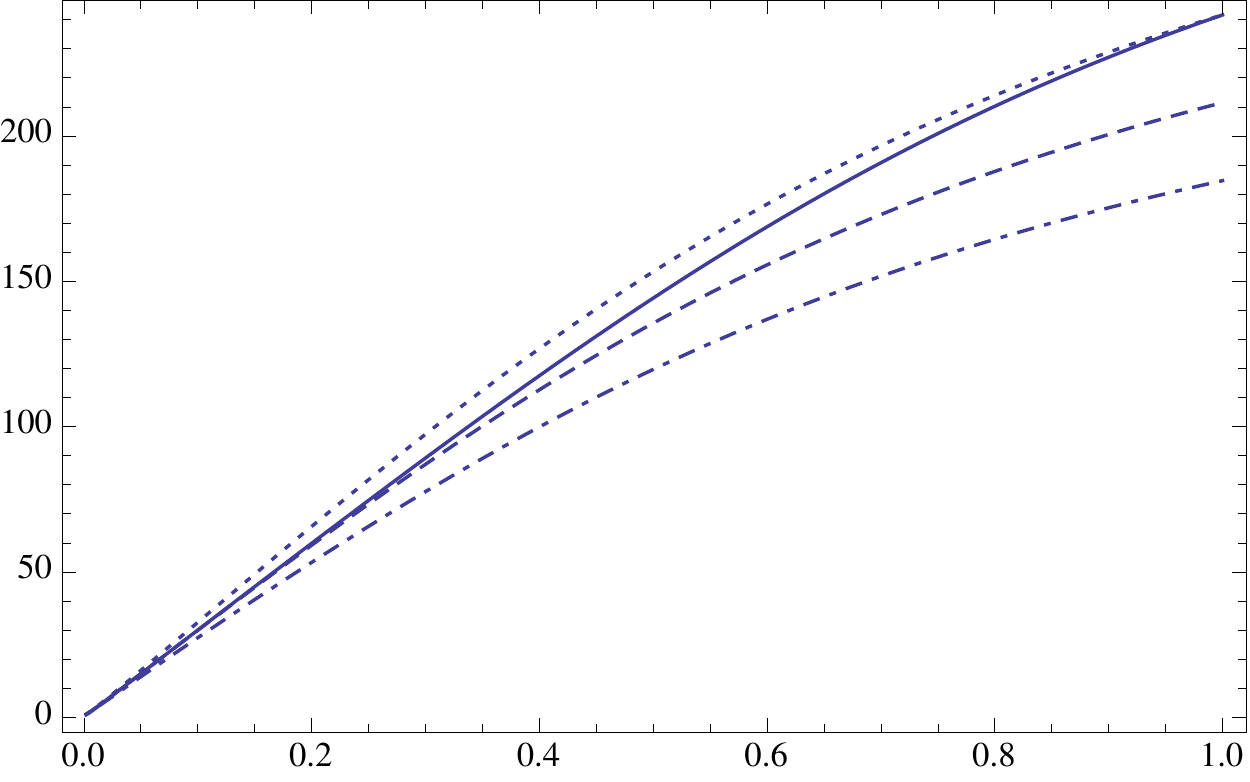}
\label{}	
\caption{The matter perturbation $\delta_{m}$ for $c=s=0$ with $\nu =0.1$ and $\mu$-values as in FIG.~6.
(dotted curve for $\mu = 1.05$, dashed curve for $\mu = 1.0$, dot-dashed curve for $\mu = 0.95$).
For $\nu\lesssim 0.1$ the curves remain in the vicinity of the $\Lambda$CDM reference curve. }
\end{center}
\end{figure}

\section{Summary}
\label{summary}

In this paper we allowed for a competition between a generalized Chaplygin gas (GCG) and a scalar field (SF) to find out
which of these is preferred by the data as the dynamically dominating component of  the cosmic substratum.
Both GCG and SF based models have found ample use separately to describe the dark sector with the intention to ``dynamize" the cosmological  constant.
To assess the background dynamics we used SNIa data from the JLA sample and $H(z)$ data both from the differential age of old galaxies that have evolved passively and from BAO.
Our combined CGSF model is characterized by 5 free parameters, where we used the CPL parametrization $\omega_{s} = \omega_{0} + \omega_{1}\left(1-a\right)$ for the scalar field. We fixed $\omega_{0}$ and the baryon fraction. The general analysis with all 5 parameters left free indicates degeneracies in the parameter space.
A two-step analysis in which the Hubble constant is fixed to its best-fit values by a suitable marginalization procedure
results in a viable two-component model of the dark sector.
The limiting cases of pure Chaplygin-gas and pure scalar-field dark sectors are consistently recovered.
A robust picture is obtained in terms of the one-particle distribution functions for each of the parameters.
The observation seem to prefer a SF fraction of more than 60\%, leaving less than 40\% for the GCG.
Moreover we found a GCG parameter $\alpha$ close to zero and a present EoS parameter for the GCG between
$-0.6$ and $-0.7$. The CPL parameter $\omega_{1}$ is positive and of the order of one.

An approximate perturbation analysis reveals that for fluctuations of the GCG energy density that are of the same order as fluctuations of the matter density the growth rate of the latter remains close to the result for the $\Lambda$CDM model. Since the GCG component accounts for properties that are ascribed to CDM, this does not come as a surprise.
For $\mu \lesssim 1$ the matter growth is reduced which corresponds to less structure formation, a possibly desired feature, given the overproduction of small-scale structure in the standard model.
On the other hand, fluctuations of the SF energy density of the order of the matter fluctuations result in unacceptable strong deviations from the standard model.
Perturbations in the SF component have to be much smaller than perturbations in the GCG component. This is consistent with the result that SF fluctuations are negligible on sub-horizon scales \cite{Staro98}.
In conclusion, while the SF dominates the homogeneous and isotropic background dynamics, it is the GCG component which governs the matter growth.
A more advanced analysis with a detailed gauge-invariant perturbation theory will be the subject of a forthcoming paper.

\acknowledgments{This work was supported by the ``Comisi\'{o}n
Nacional de Ciencias y Tecnolog\'{\i}a" (Chile) through the
FONDECYT Grant   No. 1130628 (R.H.).
J.C.F and W.Z acknowledge support by ``FONDECYT-Concurso incentivo a la
Cooperaci\'{o}n Internacional" No. 1130628 as well as by CNPq (Brazil) and FAPES
(Brazil). Sadly, shortly after this paper was started Prof. Sergio del Campo, unexpectedly, passed away. JF,RH and WZ
dedicate this paper to his Memory.}



\begin{thebibliography}{99}

\bibitem{planck}
P.A.R. Ade et al. [Planck Collaboration],
\textit{Planck 2015 results. XIII. Cosmological parameters},
arXiv:1502.01589.

\bibitem{buchert15}
  T.~Buchert, A.A.~Coley, H.~Kleinert, B.F.~Roukema and D.L. Wiltshire, \textit{Observational challenges for the standard FLRW model}, Int.J.Mod.Phys. D \textbf{25}, 1630007 (2016);
  arXiv:1512.03313.

  \bibitem{Riess} A.G. Riess et al., \textit{Observational evidence from supernovae for an accelerating universe and a cosmological constant},
Astron. J. {\bf 116}, 1009 (1998).
\bibitem{Schmidt} B. Schmidt et al., \textit{The High-Z Supernova Search: Measuring Cosmic Deceleration and Global Curvature of the Universe Using Type Ia Supernovae},
Astrophys. J {\bf 507}, 46 (1998).
\bibitem{Perlm} S. Perlmutter et al., \textit{Measurements of $\Omega$ and $\Lambda$ from 42 High-Redshift Supernovae},
Astrophys. J. {\bf 517}, 565 (1999).

\bibitem{wetterich}
C. Wetterich,
\textit{Cosmology and the fate of dilatation symmetry},
Nucl. Phys. B \textbf{302}, 668 (1988);  \textit{The cosmon model for an asymptotically vanishing time-dependent cosmological``constant''},
Astron. Astrophys. \textbf{301}, 321 (1995).

\bibitem{Ratra}  B. Ratra, and P.J.E. Peebles, \textit{Cosmological consequences of a rolling homogeneous scalar field},
Phys. Rev. D {\bf 37}, 3406 (1988).

\bibitem{CDS} R.R. Caldwell, R. Dave, and P.J. Steinhardt, \textit{Cosmological imprint of an energy component with general equation of state},
Phys. Rev. Lett. {\bf 80}, 1582 (1998).

\bibitem{CCQ} L. Wang, R.R. Caldwell, J.P. Ostriker, and P.J. Steinhardt,  \textit{Cosmic Concordance
and Quintessence},
Astrophys. J {\bf 530}, 17 (2000).

\bibitem{PeeVi} P.J.E. Peebles and A. Vilenkin, \textit{Quintessential inflation},
Phys. Rev. D {\bf 59}, 063505 (1999).

\bibitem{Zlatev}  I. Zlatev, L. Hwang, and P.J. Steinhardt, \textit{Quintessence, cosmic coincidence, and the cosmological constant},
Phys. Rev. Lett. {\bf 82}, 896 (1999).
\bibitem{Steinh}  P.J. Steinhardt, L. Hwang, and I. Zlatev, \textit{Cosmological tracking solutions},
Phys. Rev. D {\bf 59}, 123504 (1999).

\bibitem{Martin} P. Brax and J. Martin, \textit{Robustness of quintessence},
Phys. Rev. D {\bf 61}, 103502 (2000);
P. Brax, J. Martin, and A. Riazuelo, \textit{Exhaustive study of cosmic microwave background anisotropies in quintessential scenarios},
Phys. Rev. D {\bf 62}, 103505 (2000).

\bibitem{Barreiro}
T. Barreiro, E.J. Copeland and N.J. Nunes, \textit{Quintessence arising from exponential potentials},
Phys. Rev. D {\bf 61}, 127301 (2000).

\bibitem{Luca}
L. Amendola, \textit{Coupled quintessence},
Phys. Rev. D {\bf 62}, 043511 (2000); preprint astro-ph/0011243.

\bibitem{ZPC} W. Zimdahl, D. Pav\'{o}n and L.P. Chimento, \textit{Interacting
Quintessence},
\textit{Phys.Lett.} \textbf{B521}, 133 (2001).

  \bibitem{Chaplygin}
S. Chaplygin, {\it On gas jets}. Sci. Mem. Moscow Univ. Math. Phys. \textbf{21}, 1 (1904).

\bibitem{moschella} A.Y. Kamenshchik, U. Moschella and V. Pasquier, {\it An alternative to quintessence}. Phys. Lett.
{\bf B511}, 265(2001).

\bibitem{julio} J. C. Fabris, S. V. B. Gon\c{c}alves and P. E. de Souza, {\it Mass Power Spectrum in a Universe Dominated by the Chaplygin Gas}. Gen. Rel.
Grav. \textbf{34}, 53 (2002).


\bibitem{bilic}
N. Bilic, G. B. Tupper and R. D. Viollier, {\it Unification of dark matter and dark energy: the inhomogeneous Chaplygin gas}.
Phys. Lett. \textbf{B535}, 17 (2002).

\bibitem{berto}
M.C. Bento, O. Bertolami and A.A. Sen, {\it Generalized Chaplygin gas, accelerated expansion, and dark-energy-matter unification}. Phys. Rev. {\bf D66},
043507 (2002).

\bibitem{oliver08} V. Gorini, A.Y. Kamenshchik, U. Moschella, O.F. Piattella, A.A. Starobinsky,
{\it Gauge-invariant analysis of perturbations in Chaplygin gas unified models of dark matter and dark energy}.
JCAP \textbf{0802}, 016 (2008); arXiv:0711.4242.

\bibitem{oliver09} O.F. Piattella, {\it The extreme limit of the generalized Chaplygin gas}.
JCAP \textbf{1003}, 012 (2010); arXiv:0906.4430.

\bibitem{gcgjulio1} J.C. Fabris, S.V.B. Gon\c{c}alves, H.E.S. Velten, W. Zimdahl,
{\it Matter Power Spectrum for the Generalized Chaplygin Gas Model: The Newtonian Approach}.
Phys.Rev. \textbf{D78}, 103523 (2008);arXiv:0810.4308.

\bibitem{ioav} H. Sandvik, M. Tegmark, M. Zaldarriaga and I. Waga, {\it The end of unified dark matter?}. Phys. Rev. {\bf D69}, 123524 (2004).


\bibitem{rrrr} R. R. R. Reis, I. Waga, M. O. Calvao and S. E. Joras, {\it Entropy perturbations in quartessence Chaplygin models}. Phys. Rev. {\bf  D68}, 061302 (2003).

\bibitem{VDF1} W.S. Hip\'{o}lito-Ricaldi, H. E. S. Velten and W. Zimdahl, {\it Non-adiabatic dark fluid cosmology}.
JCAP \textbf{0906} 016 (2009).

\bibitem{VDF2} W.S. Hip\'{o}lito-Ricaldi, H. E. S. Velten and W. Zimdahl, {\it Viscous dark fluid Universe: a unified model of the dark sector?}
Phys. Rev. \textbf{D82}, 063507 (2010).

\bibitem{gcgjulio2} J.C. Fabris, H.E.S. Velten, W. Zimdahl,
{\it Matter power spectrum for the generalized Chaplygin gas model: The relativistic case}.
 Phys.Rev. \textbf{D81}, 087303 (2010);
 arXiv:1001.4101.

 \bibitem{nadgcg} H.A. Borges, S. Carneiro, J. C. Fabris, W. Zimdahl,
{\it Non-adiabatic Chaplygin gas}.
Phys. Lett. \textbf{B727}, 37 (2013).

\bibitem{wands}
Y. Wang, D. Wands, L. Xu, J. De-Santiago, and A. Hojjati,
{\it Cosmological constraints on a decomposed Chaplygin gas}.
Phys. Rev. \textbf{D87}, 083503 (2013).

\bibitem{saulo14} S. Carneiro and C. Pigozzo, {\it Observational tests of non-adiabatic Chaplygin gas}.
JCAP \textbf{1410}, 060 (2014); arXiv:1407.7812

\bibitem{luciano} R.F. vom Marttens, L. Casarini, W. Zimdahl, W.S. Hip\'{o}lito-Ricaldi, D.F. Mota,
\textit{Does a generalized Chaplygin gas correctly describe the cosmological
dark sector?}
Physics of the Dark Universe 15 (2017) 114–124.

\bibitem{CPL} M. Chevallier and D. Polarski, \textit{Accelerating universes with scaling dark matter}, Int.J.Mod.Phys. D \textbf{10}, 213 (2001);
E.V. Linder, \textit{Exploring the expansion history of the universe}, Phys.Rev.Lett. \textbf{90}, 091301 (2003).

\bibitem{JLA} M. Betoule et al, {\it Improved cosmological constraints from a joint analysis of the SDSS-II and SNLS supernova samples}, Astron. Astrophys, {\bf 568} 22 (2014).


\bibitem{Ji1} R. Jimenez and A. Loeb, {\it Constraining Cosmological Parameters Based on Relative Galaxy Ages}, Astrophys.J.
{\bf573} (2002) 37.

\bibitem{Ji2} R. Jimenez, L. Verde, T. Treu and D. Stern,
{\it Constraints on the equation of sate of Dark Energy and the Hubble
Constant from Stellar Ages and the Cosmic Microwave
Background}, Astrophys. J.
{\bf 593} (2003) 622.

\bibitem{Hz} D. Stern, R. Jimenez, L. Verde, M. Kamionkowski and S. Adam
Stanford, {\it Cosmic Chronometers: Constraining the Equation of State of Dark Energy. I: H(z) Measurements},
JCAP {\bf 1002} (2010) 008.

\bibitem{Farooq} O. Farooq, D. Mania and B. Ratra, {\it Hubble parameter measurement constraints on dark energy}, Astroph. J., {\bf 764} (2013) 138.

\bibitem{moresco} M.~Moresco et al., {\it New constraints on cosmological parameters and neutrino properties using the expansion rate of the Universe to $z\sim 1.75$ }, JCAP {\bf07} (2012) 053.

    \bibitem{Zheng} X. Zheng, X. Ding, M. Biesiada, S. Cao, Z. Zhu, \textit{What are Omh$^2$(z1,z2) and Om(z1,z2) \nopagebreak diagnostics telling us in light of H(z) data?}
        arXiv:1604.07910.

        \bibitem{Staro98} A.A. Starobinsky,\textit{ How to determine an effective potential for a variable cosmological term},
        JETP Lett. 68 (1998) 757-763; Pisma Zh.Eksp.Teor.Fiz. 68 (1998) 721-726.

\end{thebibliography}
\end{document}